\documentclass[a4paper,fleqn,usenatbib]{mnras}

\usepackage{newtxtext,newtxmath}

\usepackage[T1]{fontenc}
\usepackage{ae,aecompl}

\usepackage{natbib}
\usepackage{caption}
\usepackage{placeins}
\usepackage{multirow}
\usepackage{multicol}        
\usepackage{amsmath}
\usepackage{amssymb}
\usepackage{epsfig}
\usepackage{graphicx}
\usepackage{bm}		
\usepackage{pdflscape}	
\usepackage{color}

\DeclareCaptionFormat{cont}{#1 (cont.)#2#3\par}

\newcommand{\kms}{\mbox{km s$^{-1}$}}

\newcommand{\Msun}{\mbox{$M_{\odot}$}}

\newcommand{\HI}{H\,{\sevensize I}}
\newcommand{\hii}{H\,{\sevensize II}}

\newcommand{\co}{\mbox{$^{12}$CO}}
\newcommand{\cother}{\mbox{$^{13}$CO}}

\newcommand{\degrees}{\degr}

\title[The Carina Nebula and Gum 31 molecular complex]{The Carina Nebula and Gum 31 molecular complex: II.  The distribution of the atomic gas revealed in unprecedented detail.}

\author[D. Rebolledo et al.]{David Rebolledo,$^{1,2,3}$\thanks{E-mail: dreboll3@gmail.com}
Anne J. Green,$^{1}$
Michael Burton,$^{2,4}$
Kate Brooks, $^{5}$
\newauthor Shari L. Breen, $^{1}$
B. M. Gaensler, $^{6}$
Yanett Contreras, $^{7}$ 
Catherine Braiding,$^{2}$
\newauthor Cormac Purcell, $^{8}$
\\
$^{1}$Sydney Institute for Astronomy, School of Physics, The University of Sydney, NSW 2006, Australia\\
$^{2}$School of Physics, The University of New South Wales, Sydney, NSW, 2052, Australia\\
$^{3}$Departamento de Astronom\'ia, Universidad de Chile, Santiago, Chile\\
$^{4}$Armagh Observatory and Planetarium, College Hill, Armagh, BT61 9DG, Northern Ireland, UK.\\
$^{5}$Australia Telescope National Facility, CSIRO Astronomy and Space Science, P.O. Box 76, Epping, NSW 1710, Australia \\
$^{6}$Dunlap Institute for Astronomy and Astrophysics, The University of Toronto, Toronto, ON M5S 3H4, Canada\\
$^{7}$Leiden Observatory, Leiden University, PO Box 9513, NL-2300 RA Leiden, the Netherlands\\
$^{8}$Research Centre for Astronomy, Astrophysics, and Astrophotonics, Macquarie University, NSW 2109, Australia\\
}

\date{{\bf Accepted for Publication in the Monthly Notices of the Royal Astronomical Society Journal.}}

\pubyear{2017}

\begin{document}
\label{firstpage}
\pagerange{\pageref{firstpage}--\pageref{lastpage}}
\maketitle

\begin{abstract}
We report high spatial resolution observations of the \HI\ 21cm line in the Carina Nebula and the Gum 31 region obtained with the Australia Telescope Compact Array.  The observations covered $\sim$ 12 $\deg^2$ centred on $l= 287.5\degrees,b = -1\degrees$, achieving an angular resolution of $\sim 35\arcsec$. The \HI\ map revealed complex filamentary structures across a wide range of velocities.  Several ``bubbles" are clearly identified in the Carina Nebula Complex, produced by the impact of the massive star clusters located in this region. An \HI\ absorption profile obtained towards the strong extragalactic radio source PMN J1032--5917 showed the distribution of the cold component of the atomic gas along the Galactic disk, with the Sagittarius-Carina and Perseus spiral arms clearly distinguishable.  Preliminary calculations of the optical depth and spin temperatures of the cold atomic gas show that the \HI\ line is opaque ($\tau \gtrsim$ 2) at several velocities in the Sagittarius-Carina spiral arm.  The spin temperature is $\sim100$ K in the regions with the highest optical depth, although this value might be lower for the saturated components.  The atomic mass budget of Gum 31 is $\sim35 \%$ of the total gas mass.  \HI\ self absorption features have molecular counterparts and good spatial correlation with the regions of cold dust as traced by the infrared maps.   We suggest that in Gum 31 regions of cold temperature and high density are where the atomic to molecular gas phase transition is likely to be occurring.


\end{abstract}

\begin{keywords}
galaxies: ISM --- stars: formation --- ISM: molecules, dust
\end{keywords}



\section{Introduction}\label{intro}

Since its first detection, the 21 cm spin transition of \HI\ has been utilised as an excellent tool to measure the properties of the atomic phase of the interstellar medium (ISM).  This transition is readily excited under the typical physical conditions prevailing in the cold and warm ISM, resulting in a ubiquitous distribution of atomic hydrogen observed in the Milky Way (\citealt{2005ApJS..158..178M}), and in nearby galaxies (\citealt{1996ASPC..106....1W}; \citealt{2008AJ....136.2563W}).  The \HI\ 21 cm line has been extremely useful when studying the complex kinematics of the atomic gas in the local Universe (\citealt{2007ApJ...671..427M}; \citealt{2008AJ....136.2648D}).

\begin{figure*}
\begin{center}
\includegraphics[scale=0.6]{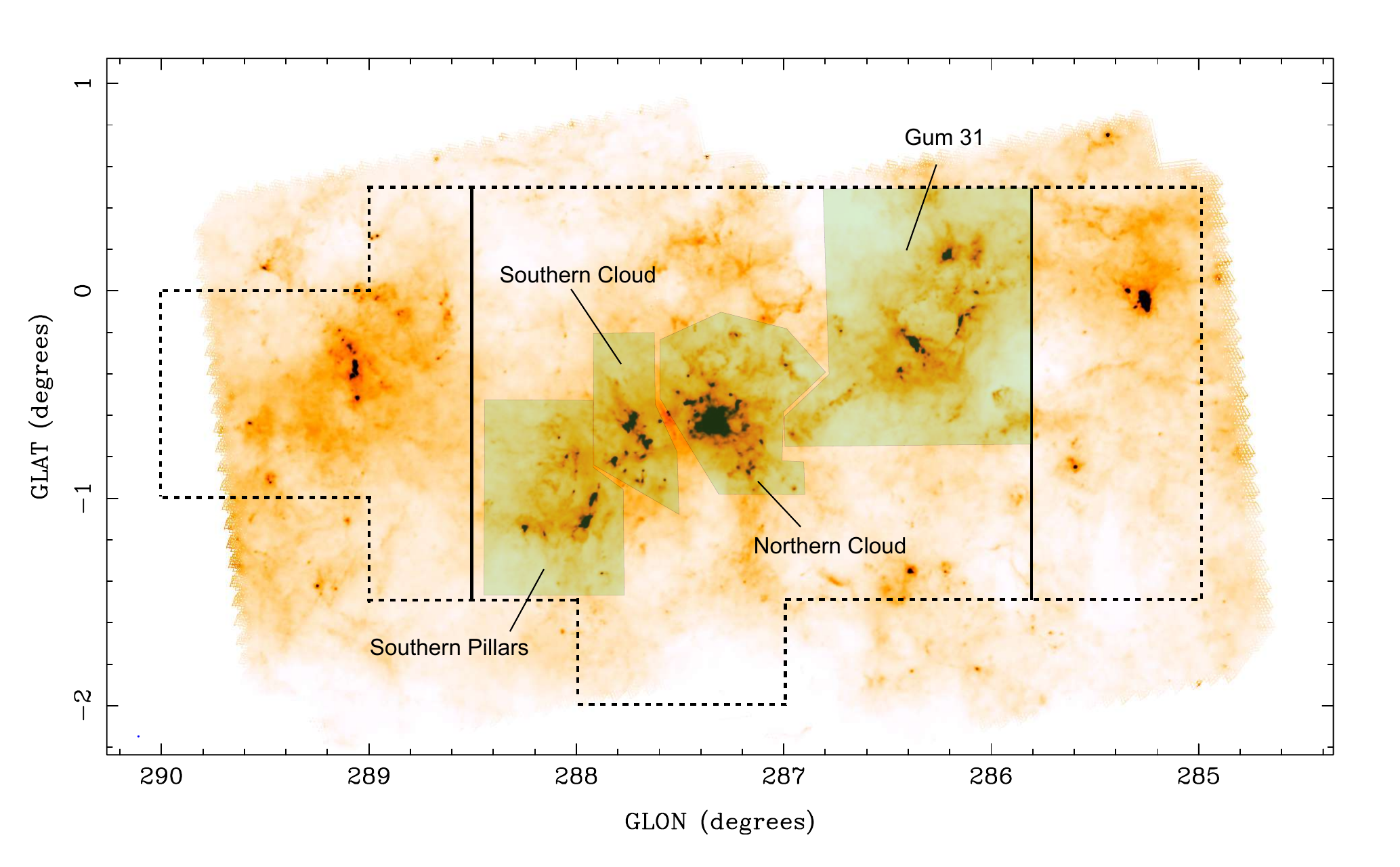}
\end{center}
\caption{{\it Herschel} 500 $\mu$m image of the CNC-Gum 31 region.  The black dashed line illustrates the region covered by the Mopra observations.  The solid black lines show the region that encloses the gas associated with the CNC and Gum 31 considered in Paper I.  Following Paper I, the Southern Pillars, the Southern Cloud, the Northern Cloud and the Gum 31 regions are shown with colour sectors.}
\label{carina.500um.obs}
\end{figure*}

However, assumptions involved in using the \HI\ 21 cm line to calculate physical properties limit the accuracy of the estimates.  For example, in order to properly measure the column density, it is necessary to know the spin temperature and the optical depth for each distinct gas cloud along the line-of-sight.  Previous analyses of emission maps have typically assumed an optically thin regime for the \HI\ line, and a single temperature for the emitting gas.  This is because, in general, decoupling the cold and warm components from the observed \HI\ line profile is challenging (\citealt{2014ApJ...781L..41M}; \citealt{2015ApJ...804...89M}).  Mapping absorption profiles against bright background continuum sources has been used as a technique to isolate the cold component, but this approach is limited by the relatively small number density of bright continuum background sources (\citealt{2000ApJ...536..756D}; \citealt{2003ApJS..145..329H}; \citealt{2003ApJ...586.1067H}).  Observations of \HI\ self-absorption (HISA) features produced by cold atomic gas in front of a warmer diffuse component have also been used to trace the cold gas (\citealt{2000ApJ...540..851G}).  However, there are difficulties in estimating the background emission spectrum that would be directly behind the HISA feature, and beam dilution effects due to the low angular resolution of current panoramic \HI\ surveys have prevented the use of this approach over larger regions.

The conversion of \HI\ into H$_{2}$ in molecular clouds is an important step in the complex process of star formation.  High resolution observations of the different gas phases in star-forming clouds are essential for studying the physical conditions prevailing in these regions.  The goal of this paper is to provide high resolution observations of the atomic gas, which combined with already available maps of the molecular gas, will help to advance the understanding of the transition of hydrogen gas from the atomic to the molecular phase in regions of active star formation (\citealt{2009ApJ...693..216K}; \citealt{2010ApJ...709..308M}; \citealt{2015ApJ...809...56L}).

The Carina Nebula Complex (CNC) and its close neighbour, the \hii\ region Gum 31, represent our nearest domains of vigorous star formation, full of massive stars and a host of new protostars (\citealt{2006MNRAS.367..763S}; \citealt{2010MNRAS.406..952S}).  The results presented here are from a major survey to map different phases of the ISM across the entire CNC-Gum 31 region using the Australia Telescope Compact Array (ATCA). This project will make available to the science community high quality radio images that will be used to understand the ecology of the ISM in the CNC-Gum31 region.  

\begin{figure*}
\begin{center}
\includegraphics[scale=0.3]{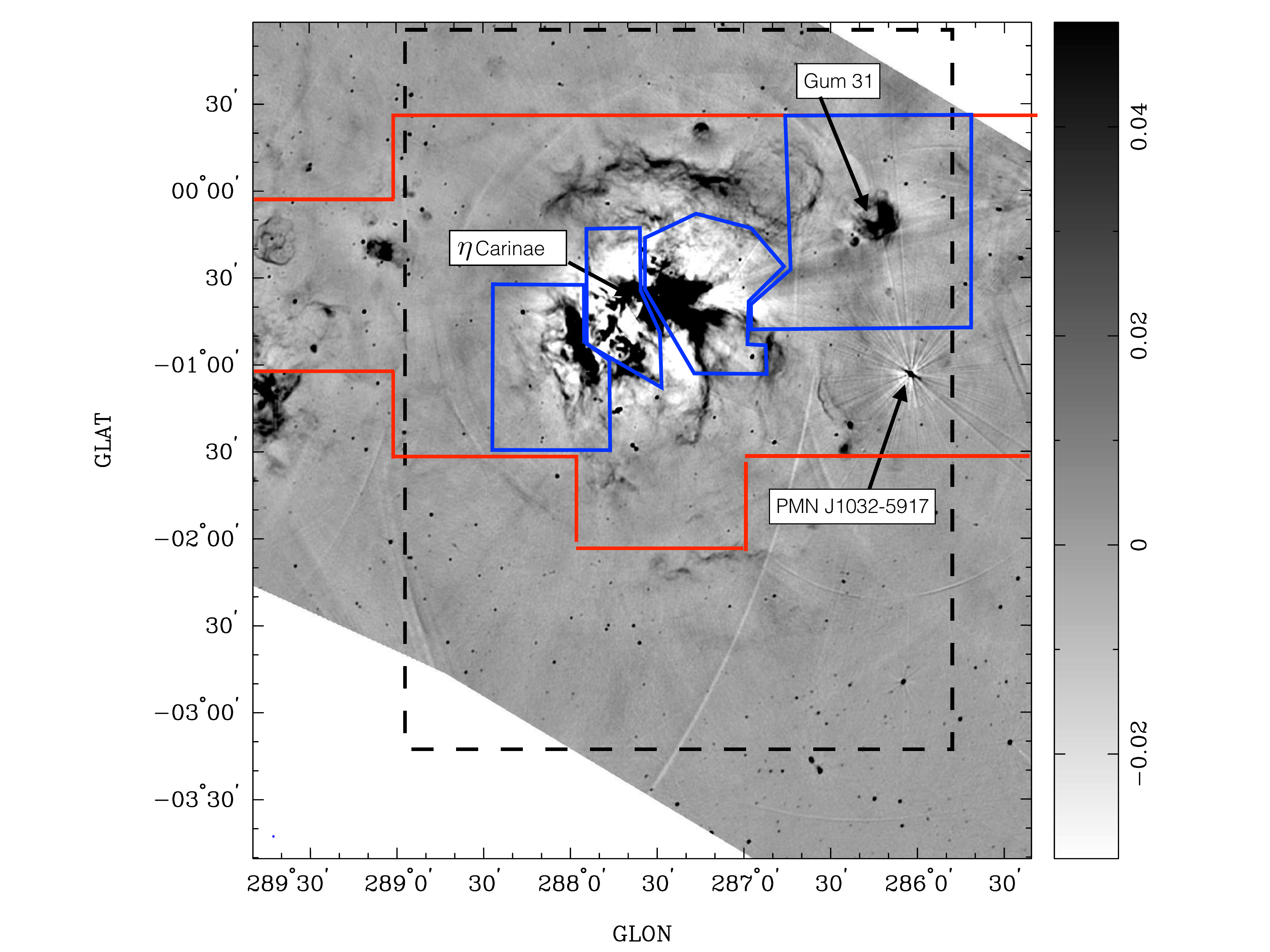}
\end{center}
\caption{Image of the 0.835 GHz continuum emission of the CNC-Gum31 complex obtained with the Molonglo Observatory Synthesis Telescope.  The bar is in units of Jy $\mathrm{beam}^{-1}$.  This image reflects the complex structure of the continuum emission across the nebula.  The black dashed box shows the region covered by our ATCA observations.  The positions of $\eta$ Carinae, Gum 31 and the strong continuum source PMN J1032-5917 are shown in the map.  Blue solid line boxes show the different regions identified in Figure \ref{carina.500um.obs}, and the red line shows the region covered by the Mopra observations.}
\label{cont_most}
\end{figure*}

In a recent paper (\citealt{2016MNRAS.456.2406R}, hereafter Paper I), we presented the $\co$ and $\cother$ maps of the CNC-Gum 31 molecular complex obtained with the Mopra telescope.  Using the CO maps, we estimated the molecular gas column density distribution across the complex.  We calculated the total gas column density and dust temperatures by fitting a grey body function to the far-infrared spectral energy distribution (SED) at every point in the target region from Herschel maps (Hi-GAL, \citealt{2010PASP..122..314M}).  Figure \ref{carina.500um.obs}, shows the regions defined in Paper I and used in this paper:  the Southern Pillars, the Southern Cloud, the Northern Cloud and the Gum 31 regions.   

This paper is the second in a series of studies that investigate the physical properties of the different gas tracers in the CNC-Gum 31 region.  We present high resolution observations of the \HI\ 21cm line that trace the atomic gas component of the ISM.  The paper is organised as follows:  In Section \ref{obs} we detail the main characteristics of the observations and the procedure to calibrate and image the data.  In Section \ref{results} we show the complex structure of the atomic gas.  We have used an absorption profile towards a strong radio continuum source in order to estimate the spin temperature and optical depth of the line.  Calculation of optical depth using self absorption profiles is also presented in Section \ref{results}.  In Section \ref{discuss} we discuss the main results and the pathway for a more complex analysis of the \HI\ 21 cm line to be published in an upcoming paper.  Section \ref{summary} summarises the conclusions.

\section{Data}\label{obs}

\subsection{Observations}   
The \HI\ 21cm line observations reported here were obtained with the ATCA synthesis imaging telescope, and cover the region $285.8\degrees \lesssim l \lesssim 289\degrees$ and $-3.0\degrees \lesssim b \lesssim 1.0\degrees$.  Figure \ref{cont_most} shows a map of the 0.835 GHz continuum emission in the CNC-Gum 31 complex obtained with the Molonglo Observatory Synthesis Telescope (\citealt{2007MNRAS.382..382M}).  In this figure, we illustrate the area covered by our ATCA observations, and the position of strong continuum sources such as $\eta$ Carinae, Gum 31 and PMN J1032--5917.  The area was observed as four mosaics each composed of 133 pointings.  The position of each pointing was chosen to ensure Nyquist sampling at the highest observed frequency.  The data were taken between April of 2011 and February of 2015 using 11 array configurations for each mosaic.  The Carina Nebula has emission features from the sub-arcsecond scale, including strong $\approx$2 Jy radio emission associated with the luminous star $\eta$ Car, to the degree scale. The observing strategy was chosen to minimise telescope artefacts and maximises the dynamic range of the images. The optimal set of array configurations selected was 6A, 6B, 6C, 1.5A, 1.5B, 1.5D, 750A, 750B, 750C, 750D and EW352, sampling $u$-$v$ baselines from 30.6 m to 5 km.  From 30.6 m to 795.9 m, we achieved an almost fully uniform sampling in intervals of 15.3 m. There was some duplication of the shorter baselines, but this was necessary as radio frequency interference often affected the lower part of the frequency band, and was strongest on the shorter baselines.  Substantial removal of corrupted samples was required. Hence, to preserve our sensitivity for the largest scale structures, observations with the EW352 array were undertaken. 

\begin{figure*}
\includegraphics[scale=0.45,angle=-90]{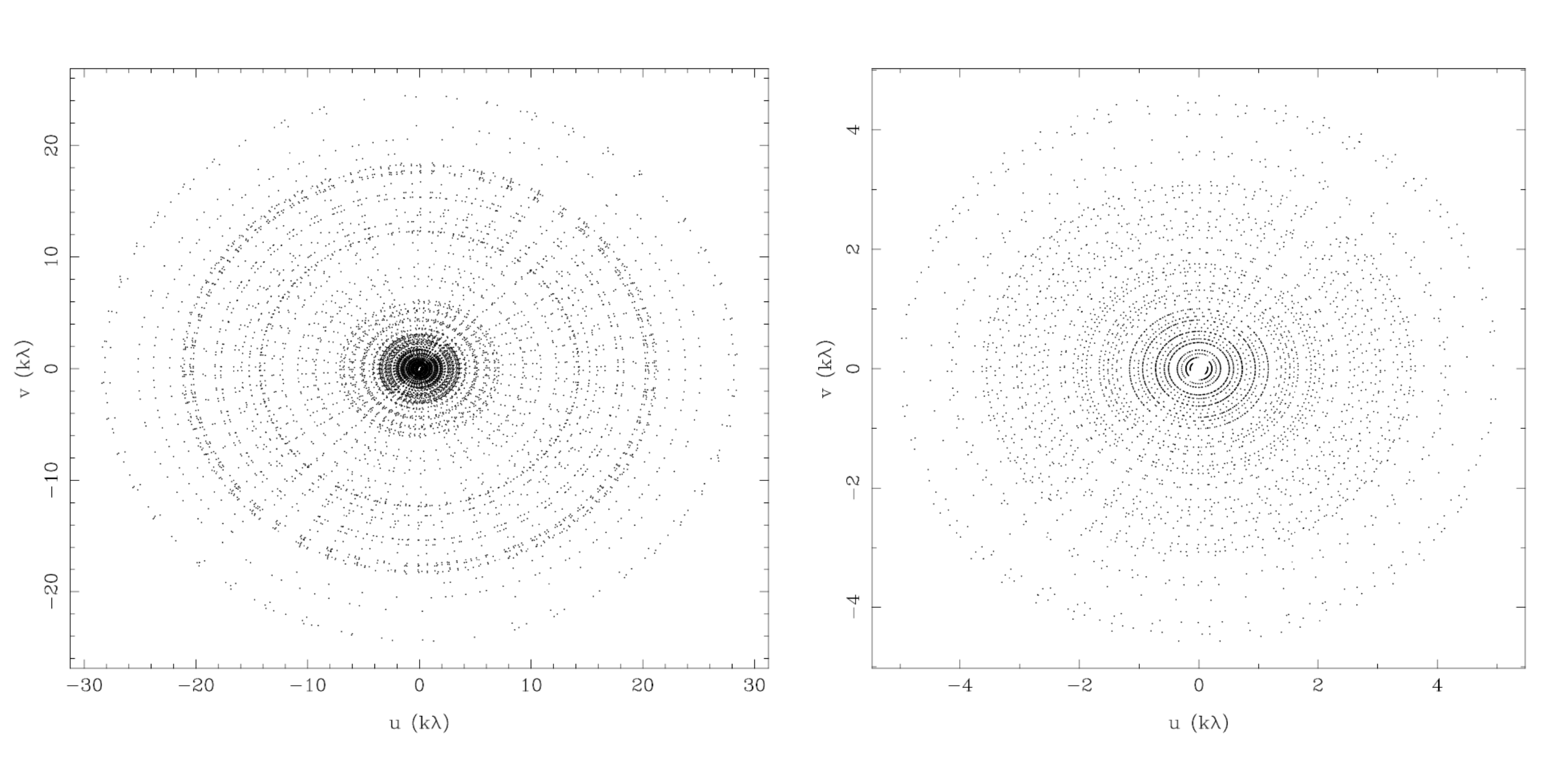} 
\caption{{\bf Left}:  The typical $u$-$v$ coverage achieved for a single  pointing using the eleven ATCA configurations. {\bf  Right}:  The $u$-$v$ coverage for the central 5 k$\lambda$ baselines.}  
\label{uvcov}
\end{figure*}

We observed each mosaic in a given array configuration for an average of 12 hours.  The resulting $u$-$v$ sampling of an individual pointing may vary slightly across the observed field.  However, our observational strategy has provided an approximately uniform $u$-$v$ coverage from pointing to pointing.  Figure \ref{uvcov} illustrates the $u$-$v$ coverage achieved  for a representative pointing.

The Compact Array Broadband Backend (CABB, \citealt{2011MNRAS.416..832W}) was used in the CFB 1M-0.5k configuration.  Four zoom bands were placed to cover 5121 channels at 0.5 kHz resolution centred at 1.42022 GHz.  This configuration provides a velocity coverage from $-$200 $\kms$ to 250 $\kms$, with a raw velocity resolution of 0.1 $\kms$. 

\subsection{Calibration}
We used standard techniques to carry out flagging and calibration with the MIRIAD data reduction package (\citealt{1995ASPC...77..433S}).  The primary flux calibrator was PKS B1934--638, which was observed at least once per observing run.  This source was used for passband and absolute flux calibration, with an assumed flux density of 14.86 Jy at 1.420 GHz.  The secondary calibrator PKS B1036--697 was observed for 3 minutes after each complete mosaic cycle, which typically took $\sim$ 20 minutes.  Once the calibration solutions were applied, we subtracted the continuum emission from the calibrated spectral cube.  This subtraction was made in the $u$-$v$ plane before imaging.

\begin{figure}
\centering
\includegraphics[scale=0.33]{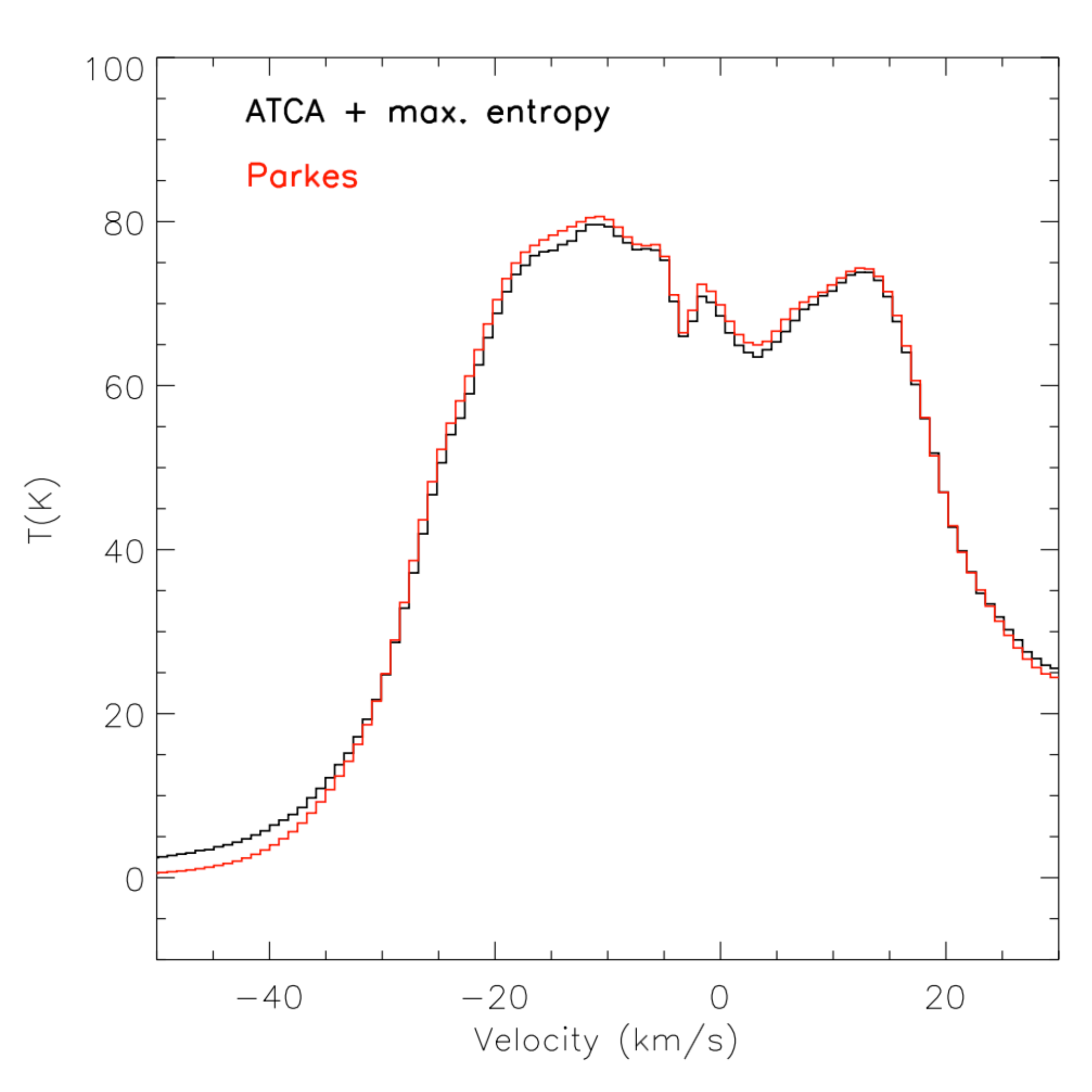}
\caption{Comparison between one \HI\ 21-cm line spectrum from this ATCA results using maximum entropy (black line) and Parkes (red line).  The profiles were obtained from a box centred at $(l,b)=(287.32\degrees, 0.18\degrees)$ with $\Delta l \times \Delta b = 1.4 \times 0.8\ \mathrm{deg}^2$.  The ATCA results have successfully recovered the total flux as traced by the Parkes profile.}  
\label{spect_comp}
\end{figure}

\subsection{Imaging}
A joint deconvolution approach was used for imaging (\citealt{1996A&AS..120..375S}).  Firstly, a dirty image for each individual pointing was generated using a standard grid and Fast Fourier Transform (FFT).  All the dirty images were then combined linearly to create a mosaiced dirty map.  The deconvolution task was divided into two steps.  Given the bright continuum emission (compact and diffuse) present in the CNC region (see Figure \ref{cont_most}), strong negative components were present in the dirty map caused by the subtraction of the continuum from the line cube before imaging.  Our preferred algorithm for deconvolving diffuse emission, applying Maximum Entropy, does not recognise negative components.  Hence, we initially identified the negative components from the mosaiced dirty map by deconvolving with the CLEAN algorithm (\citealt{1974A&AS...15..417H}; \citealt{1980A&A....89..377C}).  These negative components were then subtracted from the dirty image.  The residual map  was then deconvolved using the Maximum Entropy algorithm (\citealt{1999ASPC..180..151C}).  In order to recover the total flux for the \HI\ maps, we used the Parkes \HI\ cube from the Southern Galactic Plane Survey (SGPS, \citealt{2005ApJS..158..178M}).  The Parkes \HI\ data were input as the initial solution for the process.  This algorithm forces the deconvolved image towards the values in the single dish image for the spatial frequencies where there is no interferometric data.  Finally, both negative and positive components were restored using a Gaussian synthesized beam of major axis, minor axis, and position angle of 37.8\arcsec, 30.5\arcsec\ and $-1.47$\degrees\ respectively.  The resulting \HI\ map has a velocity resolution of 0.5 $\kms$, and noise sensitivity in a single channel of 4.7 K.  The data will be available to the community in the website of the project\footnote{http://www.physics.usyd.edu.au/sifa/carparcs}.

\subsection{Comparison with Parkes map}\label{comp_parkes}
Figure \ref{spect_comp} shows a comparison between the \HI\ profiles from our data and Parkes (\citealt{2005ApJS..158..178M}).  The spectra were obtained from a box centred at $(l,b)=(287.32\degrees, 0.18\degrees)$ with $\Delta l \times \Delta b = 1.4 \times 0.8\ \mathrm{deg}^2$.  Because this region is away from the strong continuum sources in the CNC, the emission features are well recovered.  For this comparison, the ATCA image has been smoothed and regridded to the Parkes resolution and pixel size respectively.  Figure \ref{spect_comp} provides evidence that the interferometric \HI\ map successfully recovers the total flux as traced by the Parkes map.    

\subsection{Distribution of the atomic gas emission in CNC-Gum31}\label{hi_dist}

\begin{figure*}
\centering
\includegraphics[scale=0.72]{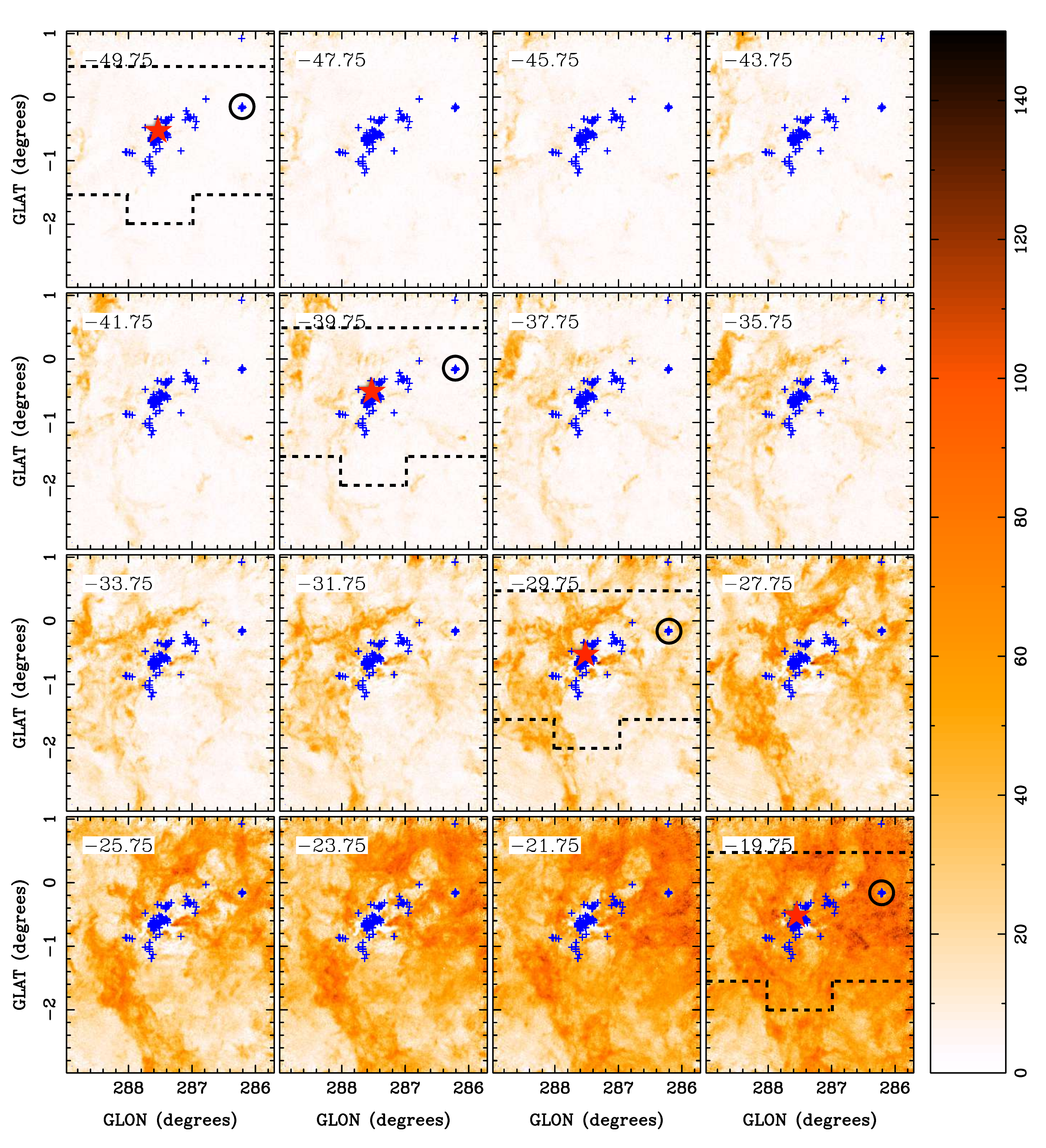}
\caption{Channel map of the \HI\ 21-cm emission line of the CNC-Gum 31 molecular complex.  The colour bar is in K, and each panel has a velocity range of 2 $\kms$.  The blue crosses show the positions of the members of the massive star clusters.  The red star shows the position of $\eta$ Carinae, and the black circle identifies the NGC 3324 star cluster associated with the \hii\ region Gum 31.  The black dashed line shows the region covered by the Mopra observations as in Figure \ref{carina.500um.obs}.}  
\label{channel1}
\end{figure*}

\begin{figure*}
\includegraphics[scale=0.72]{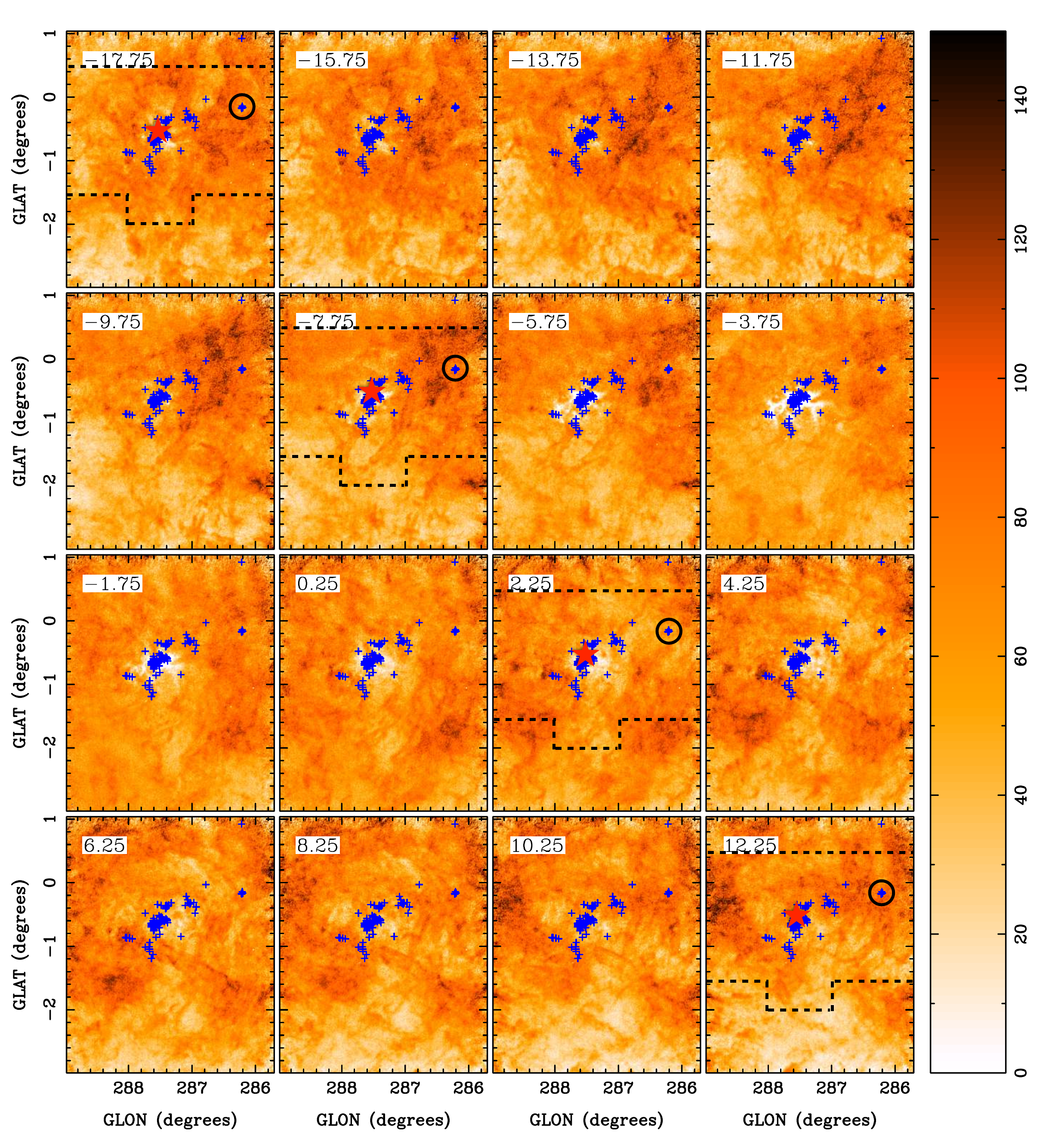}
\contcaption{}  
\label{channel2}
\end{figure*}

\begin{figure*}
\includegraphics[scale=0.72]{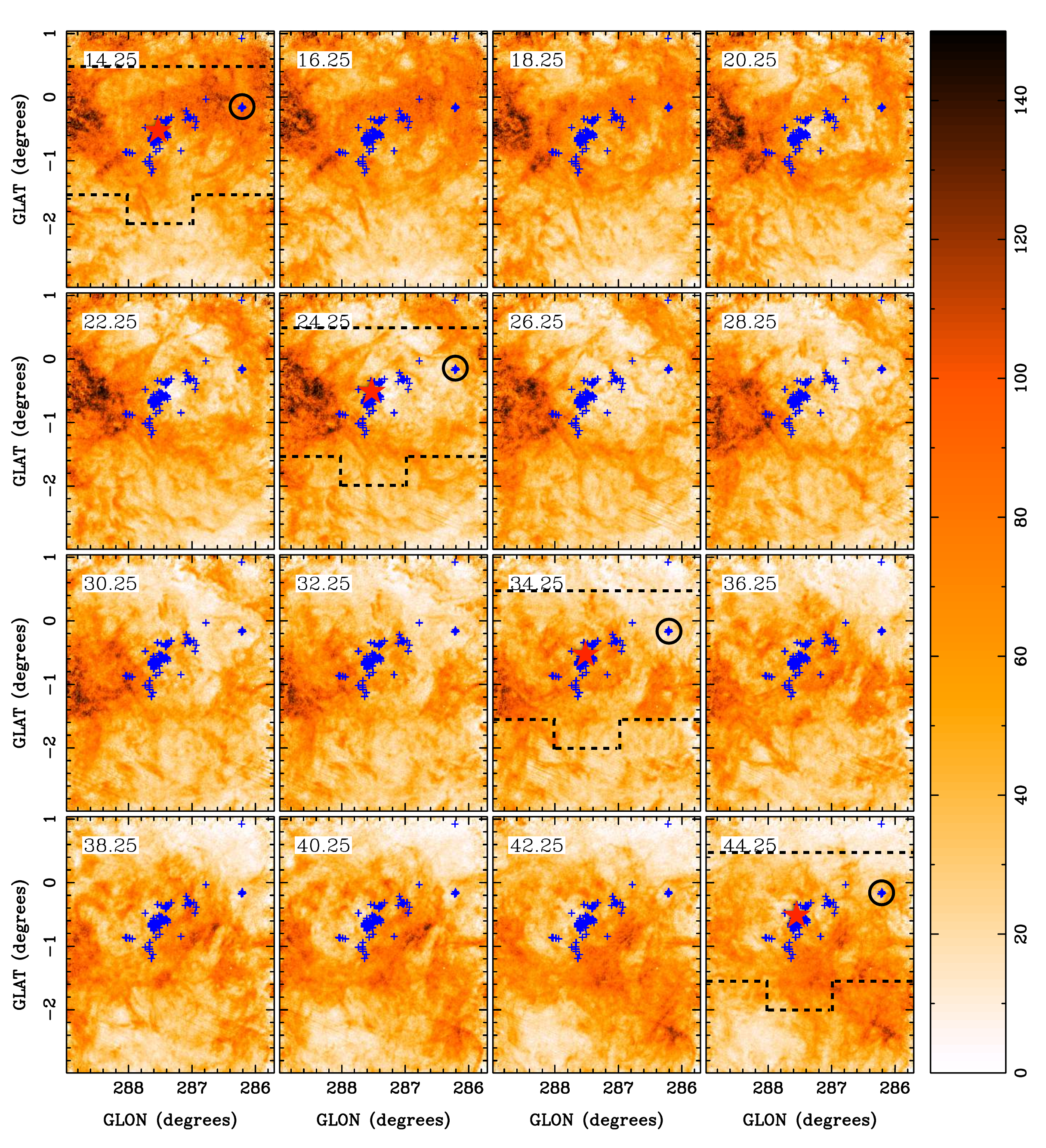}
\contcaption{}  
\label{channel3}
\end{figure*}

\begin{figure*}
\includegraphics[scale=0.72]{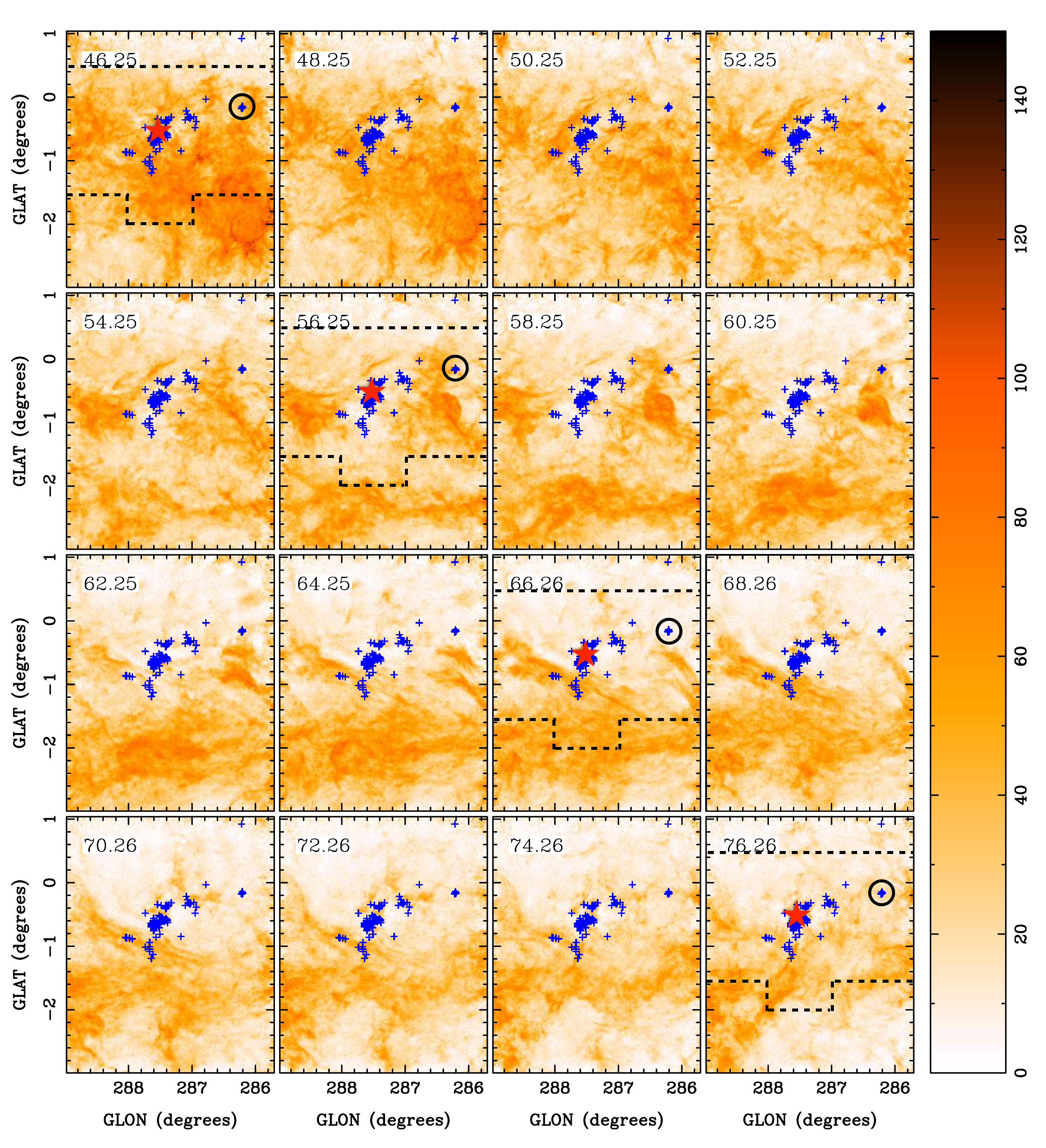}
\contcaption{}  
\label{channel4}
\end{figure*}

\begin{figure*}
\includegraphics[scale=0.72]{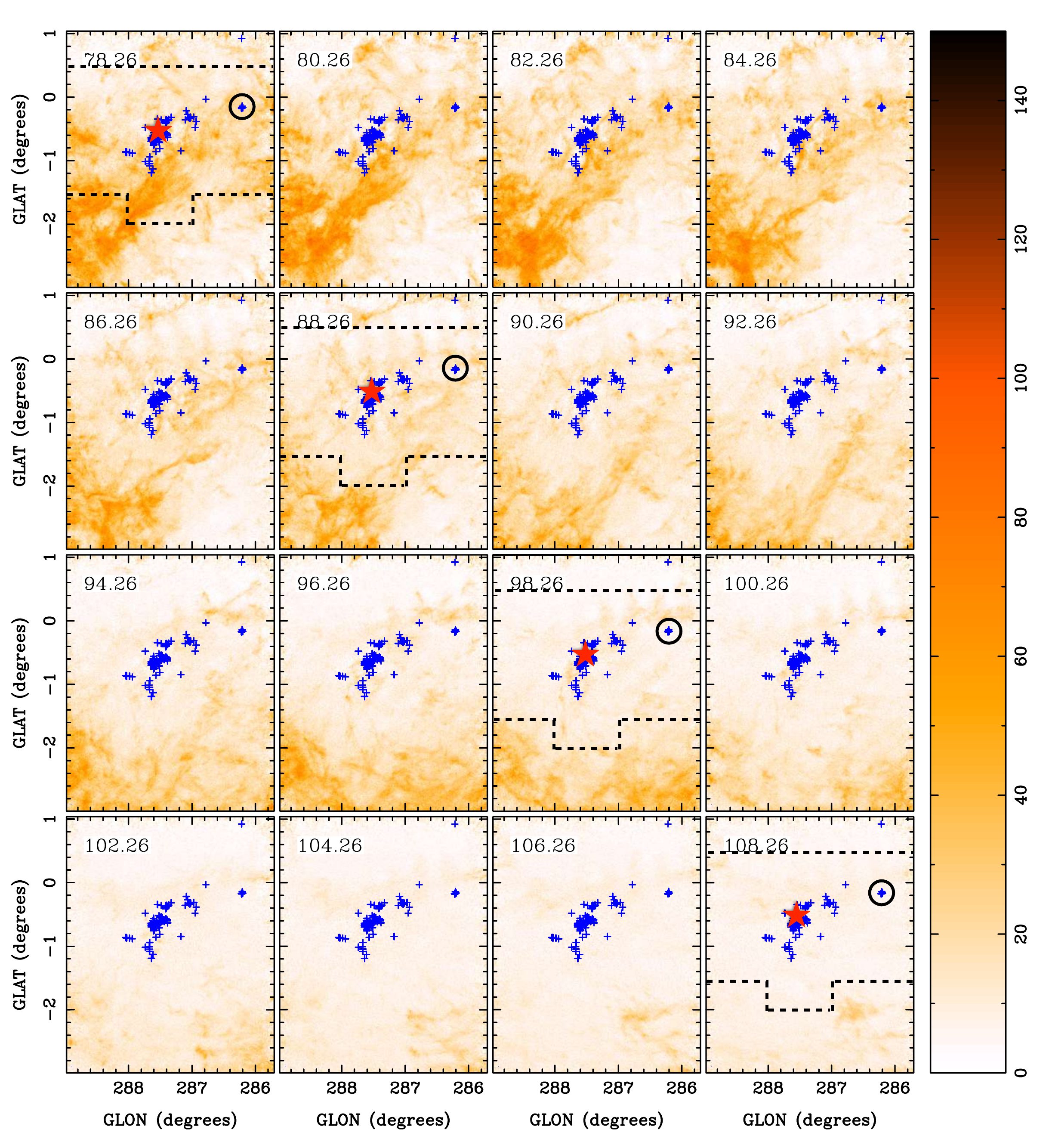}
\contcaption{}  
\label{channel5}
\end{figure*}

\begin{figure*}
\includegraphics[scale=0.72]{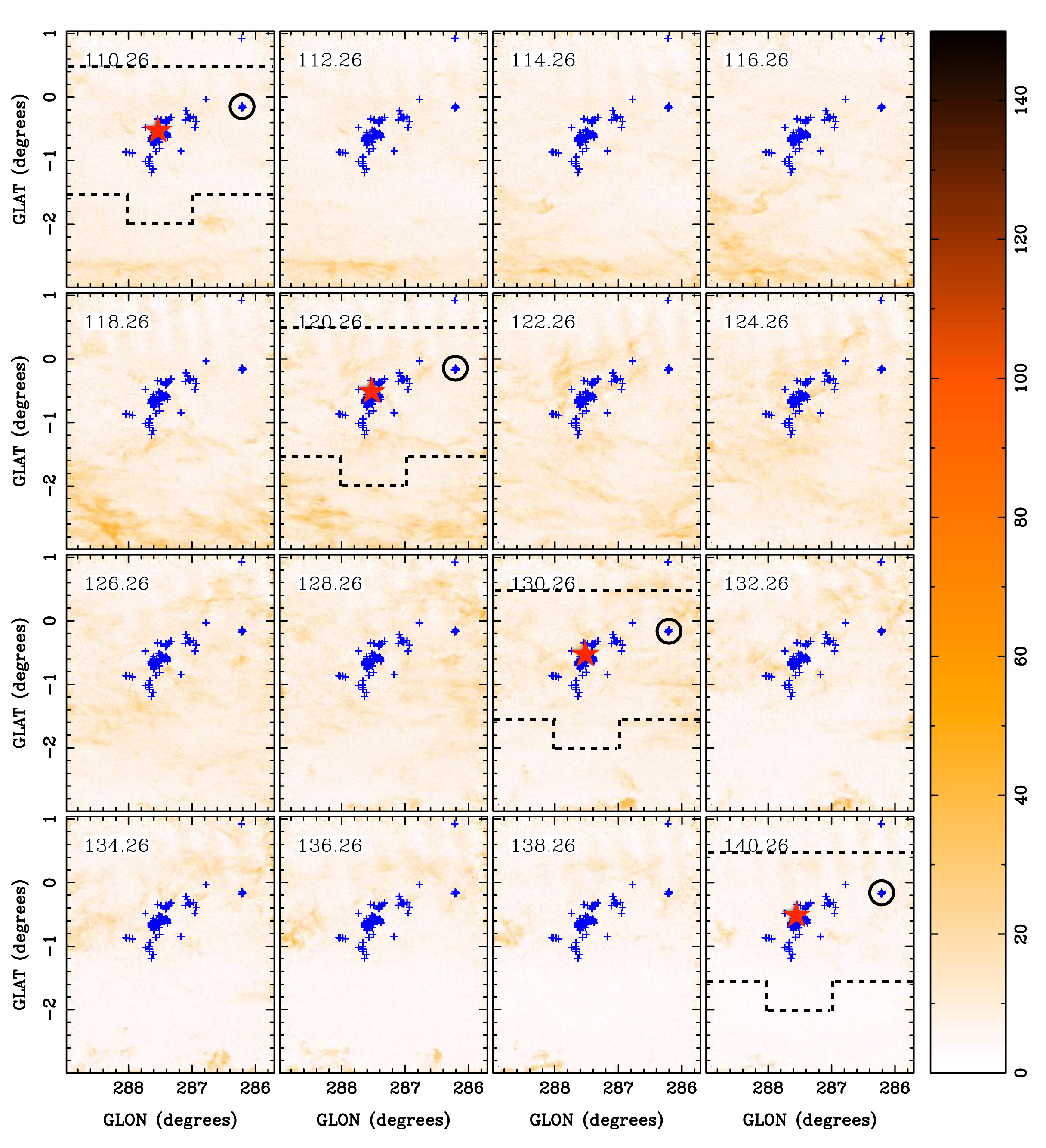}
\contcaption{}  
\label{channel6}
\end{figure*}

In Figure \ref{channel1} the resulting \HI\ channel maps of the CNC-Gum 31 region are shown.  We have averaged over 4 channels to give a velocity resolution of 2 $\kms$.  For display purposes only, the positions of the massive stars located in the complex have been marked (\citealt{2006MNRAS.367..763S}; \citealt{2001A&A...371..107C}), highlighting the position of $\eta$ Carinae.  Notice that most of the massive stars are located in the sub-regions associated with the CNC.  The star cluster NGC 3324 is the only cluster in the Gum 31 region.  We start detecting emission at about $-50\ \kms$, which corresponds to clouds located in the foreground of the line-of-sight towards the CNC.  From $-20\ \kms$, the Sagittarius-Carina spiral arm dominates the emission in the observed area.  Using the four-arm model of the Galaxy from \citet{2014AJ....148....5V}, we showed in Paper I the position-velocity (PV) diagram of the spiral arms with respect to the Local Standard of Rest frame.  According to this model, the Sagittarius-Carina spiral arm extends from $ -20\ \kms$ to $20\ \kms$ over the longitude window covered by our observations.  A complex structure of interlaced filaments and cavities is clearly seen across all velocities.  Given the superposition of emission along each line-of-sight in this region, it is intrinsically difficult to properly identify discrete structures in the \HI\ maps.  However, we can still identify certain features, especially in regions away from the Galactic mid-plane.

\section{Results}\label{results}

For gas with velocities $-35\ \kms$ to $ -25\ \kms$ we identify material that is kinematically and spatially connected.  This structure comprises a bridge of gas extending from $b=-3 \degrees$ and $l=287.7\degrees$ towards the central part of the nebula located at $b=-0.5\degrees$.   The region in this bridge located at the central part of the CNC seems to be associated with the location of the Southern Pillars as observed in the Mopra CO images (see Figure \ref{hi_co}).  

\begin{figure*}
\includegraphics[scale=0.3]{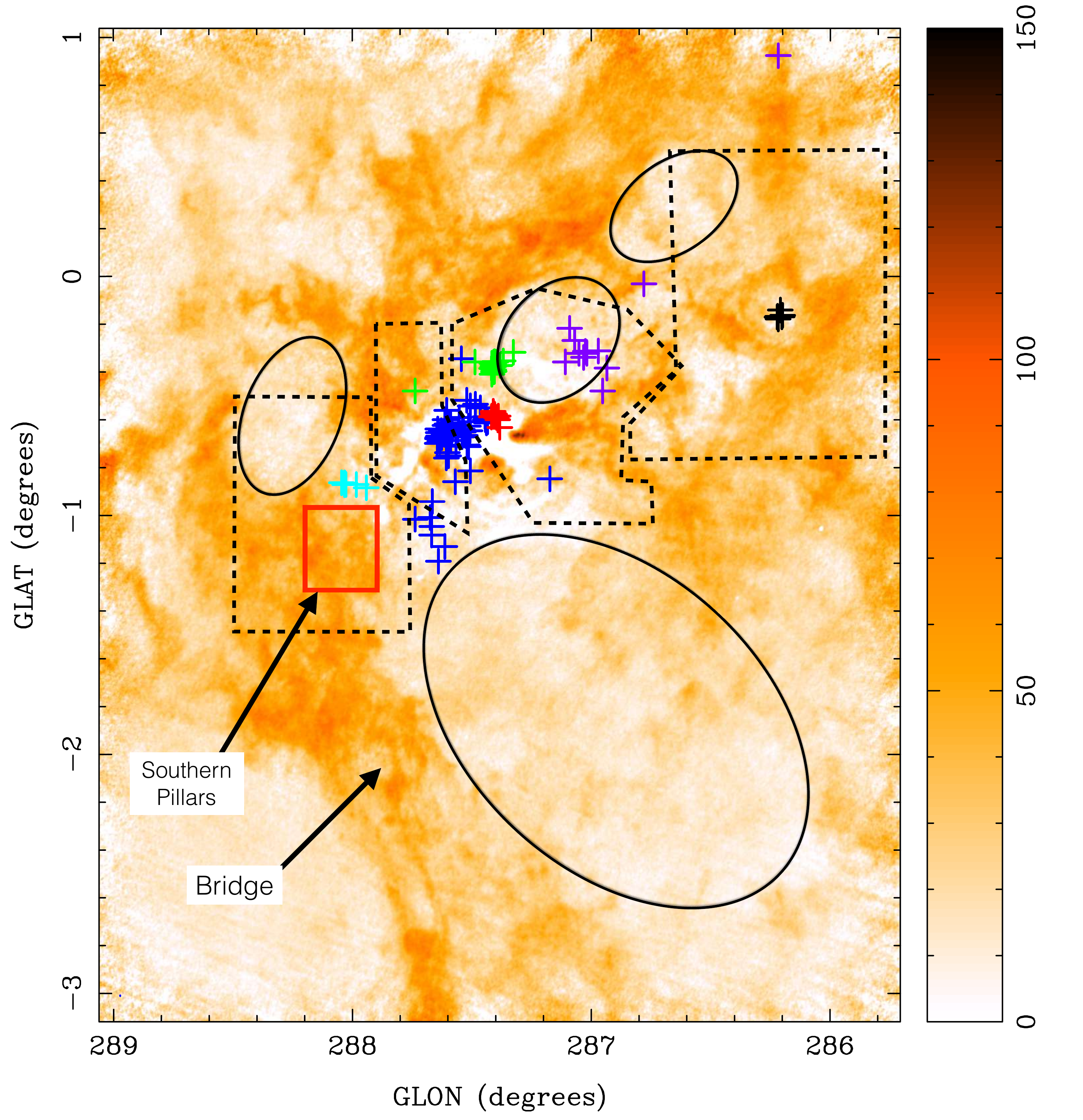}
\caption{\HI\ averaged intensity map over the velocity range $-30\ \kms$ to $-22\ \kms$.  As in Figure \ref{channel1}, crosses show the positions of the members of the massive star clusters.  The blue crosses show the Trumpler 16 and Collinder 228 stellar clusters, red crosses show Trumpler 14, green show Trumpler 15, violet shows Bochum 10 and cyan crosses show Bochum 11 cluster, all of them associated with the CNC and listed in \citet{2006MNRAS.367..763S}.  The black crosses show the star member of the NGC 3324 cluster in the Gum 31 region and listed in \citet{2001A&A...371..107C}.  The black ellipses illustrate the cavities or ``bubbles" that are distinguishable in this velocity range.  Black dashed line boxes show the different sub-regions identified in Figure \ref{carina.500um.obs}.  The red box illustrates the region where the Southern Pillars is located (see Figure \ref{hi_co}).}  
\label{cavitities}
\end{figure*}

Several cavities or ``bubbles" are clearly distinguishable between velocities $-40\ \kms$ and $-22\ \kms$.  In Figure \ref{cavitities} we show the averaged intensity map over the velocity range $-30\ \kms$ to $-22\ \kms$.  Although the cavities are not simple spherical shells, we have overlaid black ellipses that highlight the location of the most prominent cavities.  The biggest ``bubble" structure is located directly below the central core of the CNC, and extends $\sim 2\degrees$ toward the position $(l,b)=(286\degrees, -3\degrees)$.  Using infrared data from the Midcourse Space Experiment (MSX) satellite, \citet{2000ApJ...532L.145S} studied the large scale structure of the CNC.  They suggested that the CNC presents a bipolar structure or ``superbubble", with lobe diameters of $\sim 40$ pc.  The polar lobe extending to negative latitudes is consistent with the cavity we find in our \HI\ data, although our observations show that the extent of this bubble could be a factor of 2 larger.  However, the polar lobe extending towards the Galactic Plane that was identified by \citet{2000ApJ...532L.145S} overlaps with two cavities we find in the same region.  Our decomposition of the shell-like structure is not unique and a single structure could also have been proposed.  The \HI\ cavity identified above the Southern Pillars does not have a clear counterpart in the MSX image from \citet{2000ApJ...532L.145S}.  This cavity was previously named GSH 288.3--0.5--28 and potentially associated with the magnetar 1E 1048.1--5937 by \citet{2005ApJ...620L..95G}.  They suggested that GSH 288.3--0.5--28 cavity has been shaped by the wind blown by the massive progenitor of 1E 1048.1--5937. 

For velocities above $-20\ \kms$, the observed structure of the atomic gas becomes dominated by the diffuse emission of the Sagittarius-Carina spiral arm.  In Paper I, gas in the velocity range from $-20\ \kms$ to 0 $\kms$ was found to be associated with the CNC-Gum 31 region.  For gas with velocities from  $-20\ \kms$ to $-5\ \kms$ the region surrounding Gum 31 has the highest brightness temperatures, usually above 100 K.  For gas with positive velocities, the highest brightness temperatures are located at the southern part of the Galactic Plane, concentrated between $-2\degrees$ and $-1\degrees$ in latitude.  Above velocities $\sim$10 $\kms$, the emission is again concentrated within $\pm 1\degrees$ of the Galactic Equator.  At velocities larger than $\sim$ 20 $\kms$, we start to detect the edge of the Sagittarius-Carina spiral arm.  There is a bright feature located between $b= -1\degrees$ to 0$\degrees$, and $l= $288 $\degrees$ to 289$\degrees$.  This appears to be a large cloud stretching from the Sagittarius-Carina arm into the inter-arm region.  Gas in the inter-arm region has velocities from $\sim 40\ \kms$ to $70\ \kms$, and there is a clear decrease in the intensity of the emission.  One remarkable feature is a 1$\degrees$ wide horizontal band centred at $b= -2 \degrees$, spanning the velocity range from 55 $\kms$ to $75\ \kms$.  This diffuse structure may be related to an inter-arm bridge of material connecting with the Perseus arm, which starts at $\sim 75\ \kms$.  Beyond $\sim 100\ \kms$, the emission is minimal, although there are some high-velocity clouds detected at negative latitudes.

\begin{figure*}
\centering
\begin{tabular}{c}
\includegraphics[scale=0.45,angle=-90]{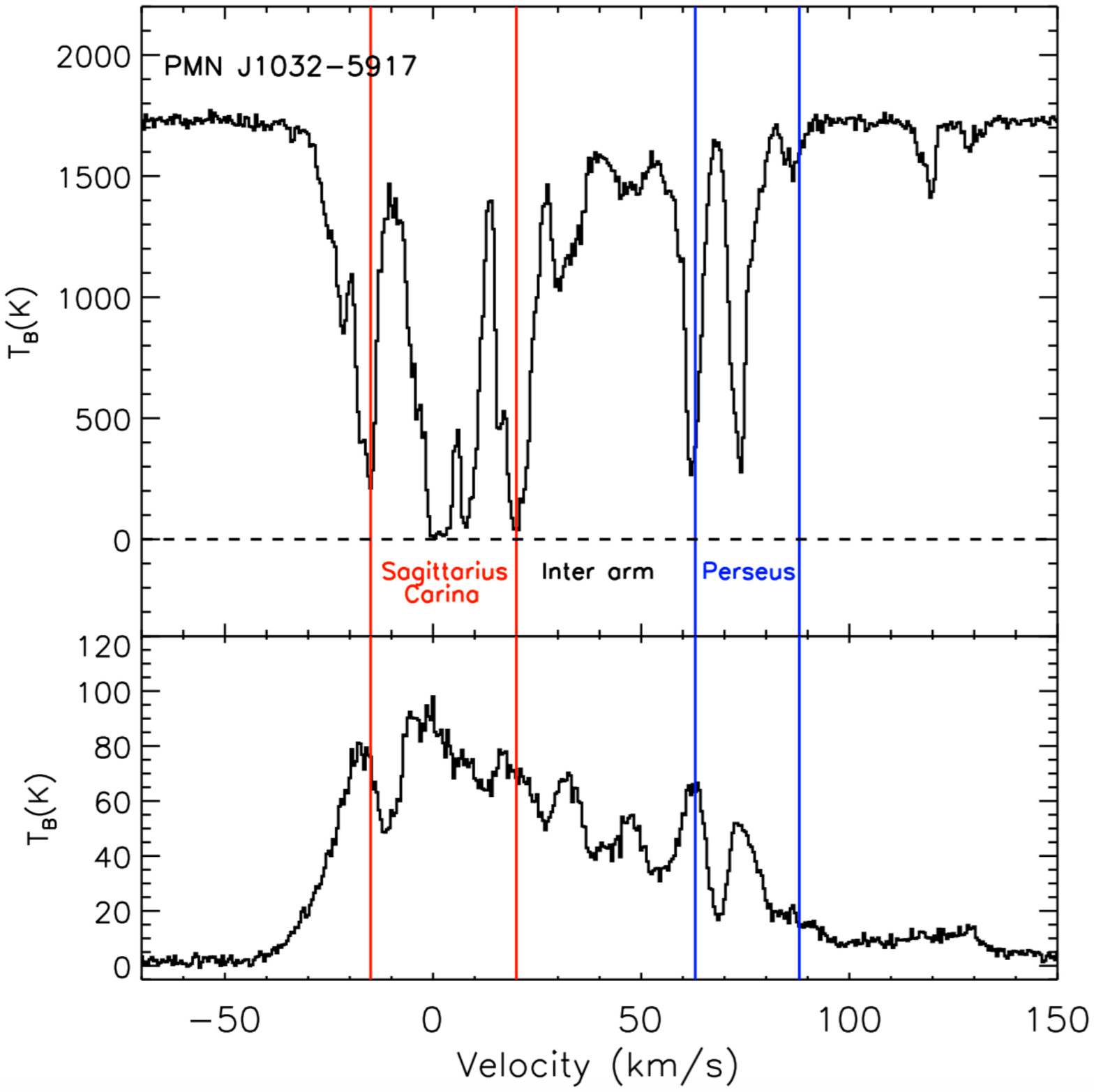}
\end{tabular}
\caption{{\bf Top}: \HI\ Absorption profile toward the strong radio source PMN J1032--5917 (\citealt{2007ApJ...663..258B}).  The absorption profile has been obtained solely from our ATCA observations with no single dish data, to maximise the removal of emission along the line-of-sight.  The absorption profile shows clearly the different components of the cold atomic gas across the Galactic disk.  The vertical lines illustrate the velocity range of gas in the spiral arms according to the model from \citet{2014AJ....148....5V}.  The red lines mark the Sagittarius-Carina arm, while the blue lines show the Perseus arm.  {\bf Bottom}:  \HI\ emission profile obtained from an annulus centred on the position of the source PMN J1032--5917.  The annulus selected has inner radius of 140$\arcsec$ and outer radius 200$\arcsec$.}  
\label{absoption_compact}
\end{figure*}

Although the \HI\ 21 cm emission maps are extremely useful for visualising the distribution of the atomic gas, they are not useful as tracers of the cold atomic gas ($T_\mathrm{S} < 100$ K).  The \HI\ brightness temperature is insensitive to the temperature when the line is optically thin, and it is proportional to the excitation temperature when it is optically thick. Hence, it is intrinsically difficult to separate the various temperature components of the \HI\ line for different opacity conditions (\citealt{2000ApJ...540..851G}; \citealt{1996ASPC..106....1W}).  On the other hand, \HI\ absorption profiles against bright continuum sources and observations of HISA features have been used to isolate the cold component (\citealt{2000ApJ...536..756D}; \citealt{2003ApJS..145..329H}; \citealt{2003ApJ...586.1067H}; \citealt{2000ApJ...540..851G}).  In the following sections, we will analyse some of the absorption features identified in the data. 

\subsection{Continuum absorption}\label{cont_abs}
\HI\ continuum absorption has been used to study the distribution and properties of the cold component of the atomic gas of the ISM (\citealt{2000ApJ...536..756D}; \citealt{2003ApJS..145..329H}; \citealt{2003ApJ...586.1067H}).  The \HI\ line will be observed in absorption if the background continuum source brightness temperature is higher than the spin temperature of the foreground \HI\ cloud.  The typical method for determining the emission brightness temperature and optical depth is the ``on-off'' method (\citealt{2000ApJ...536..756D}; \citealt{2003ApJS..145..329H}; \citealt{2003ApJ...586.1067H}; \citealt{2004ApJ...603..560S}; \citealt{2007AJ....134.2252S}), using compact extragalactic continuum sources.  In the following sections we present the absorption profiles toward the strong continuum source PMN J1032--5917 (Figure \ref{cont_most}) and the absorption profiles produced by the strong diffuse continuum emission from the \hii\ region in the CNC.

\subsubsection{Absorption profile towards the radio continuum source PMN J1032--5917}\label{radio_compact}
Figure \ref{absoption_compact} shows the \HI\ absorption profile obtained toward the radio source PMN J1032--5917 (\citealt{2007ApJ...663..258B}).  This strong extragalactic source has a peak intensity of $\sim 2.8$ Jy/beam at 1.4 GHz (\citealt{2007ApJ...663..258B}), and is located at $(l,b) = (286.04\degrees, 1.05\degrees)$.  The absorption profile has allowed us to identify the multiple components of the cold atomic gas across the Galactic disk.  The most prominent components are associated with the Sagittarius-Carina and Perseus spiral arms.  The inter-arm region shows little absorbing material.  In Figure \ref{absoption_compact} we also show the \HI\ emission profile around this compact continuum source, obtained using an annulus with internal radius 140$\arcsec$ and external radius 200$\arcsec$.  The centre of the annulus is the position of the peak-intensity pixel of PMN J1032--5917.  There is good correspondence between the cold gas components identified in the absorption profile and the intensity peaks in the emission profile.  However, the emission profile is a combination of the cold and warm components of the atomic ISM, making it more difficult to separate the different velocity components present along the line-of-sight.  In Section \ref{tau-spin} we will estimate the optical depth and the spin temperature of the \HI\ line along the line-of-sight towards PMN J1032--5917.

\subsubsection{Diffuse continuum emission from the \hii\ region in Carina}\label{cont_diffuse}
Background diffuse continuum emission can also affect the \HI\ line profile.  Weak diffuse continuum emission that is not strong enough to produce \HI\ absorption features can diminish the \HI\ line intensity (see \citealt{2015A&A...580A.112B} for example).  If an \HI\ cloud is located between the observer and the continuum source, then the observed brightness temperature along a given line of sight is (\citealt{2000ApJ...540..851G}; \citealt{1996ASPC..106....1W})

\begin{equation}\label{Tb-all}
T_\mathrm{B}=T_\mathrm{S}(1-e^{-\tau})+ T_\mathrm{C}e^{-\tau},
\end{equation}

\noindent where $T_\mathrm{S}$ and $\tau$ are the spin temperature and the optical depth of the \HI\ cloud respectively, while $T_\mathrm{C}$ is the brightness temperature of the continuum source.  Subtracting the continuum contribution from the data, the observed brightness distribution of the \HI\ line only $T_\mathrm{B,HI}$ is given by

\begin{equation}\label{Tb-contsubs}
T_\mathrm{B,HI}=(T_\mathrm{S}-T_\mathrm{C})(1-e^{-\tau}).
\end{equation}

\begin{figure*}
\centering
\begin{tabular}{c}
\includegraphics[scale=0.3]{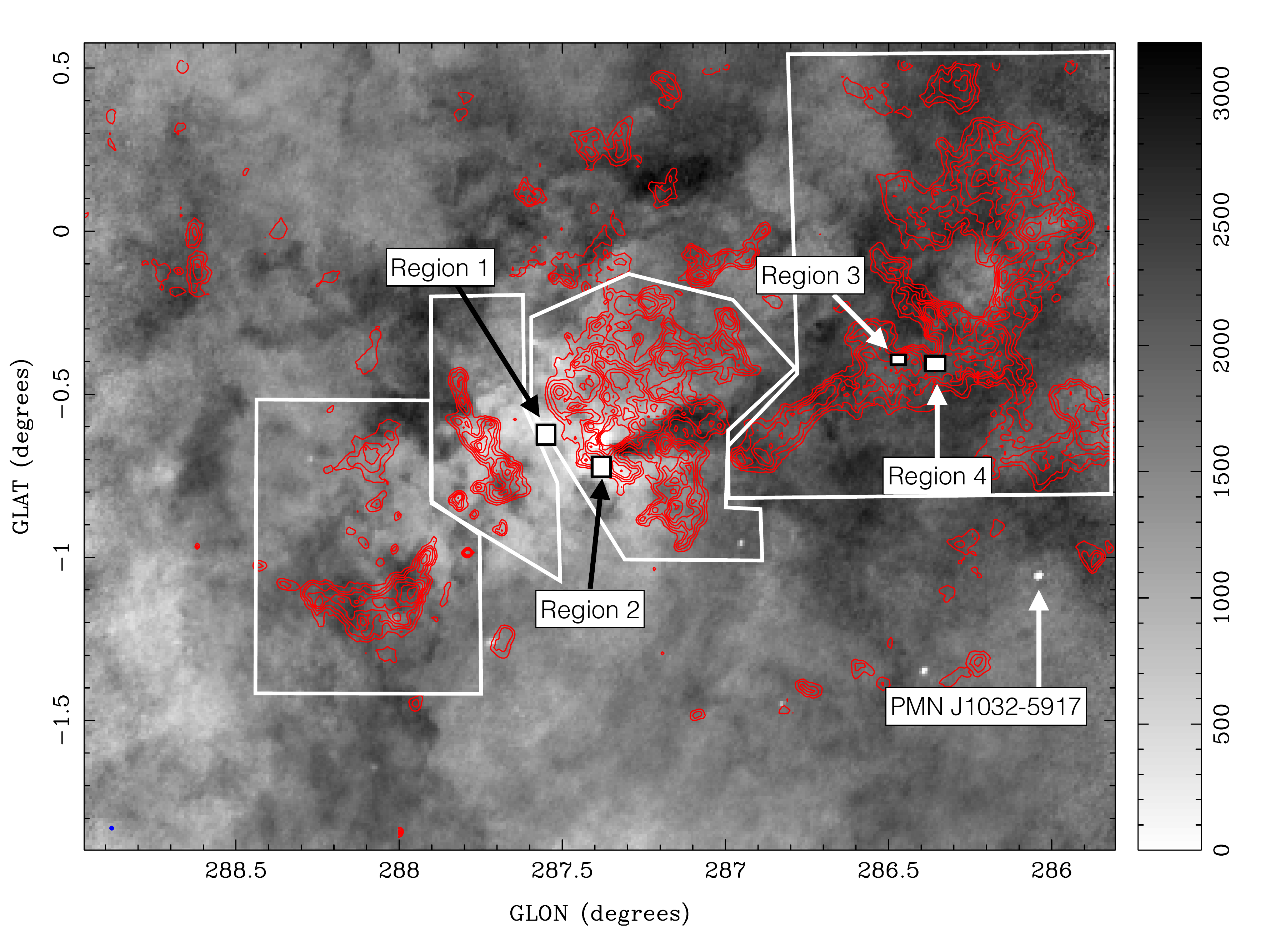}
\end{tabular}
\caption{Comparison of the \HI\ and $\co$ integrated intensity maps of the CNC-Gum31 molecular complex, over the velocity range $-45\ \kms$ to 0\ $\kms$.  The $\co$ map is from Paper I.  The color bar is in units of K $\kms$.  The greyscale shows the \HI\, while the red contours plot the $\co$ emission.  Contours are spaced by $n^2$ K $\kms$, with $n=2, 3, 4, 5, 6, 7, 8$.  The white filled boxes are the regions used to extract the \HI\ absorption spectra shown in Figures \ref{spect_abs} and \ref{spect_abs_hisa}.  The white solid lines show the sub-regions identified in Figure \ref{carina.500um.obs}.}  
\label{hi_co}
\end{figure*}

Figure \ref{hi_co} shows the \HI\ integrated intensity map of the central part of the CNC, centred on $(l,b)=(287.57\degrees, -0.75\degrees)$, and with $\Delta l=1.6\degrees \times\ \Delta b= 1.23\degrees$.  We have integrated the spectral line over the velocities from $-45\ \kms$ to $0\ \kms$, which corresponds to the range for gas associated with the CNC (Paper I).  The effect of the background diffuse continuum emission is clearly seen on the distribution of the observed \HI\ integrated intensity.  We observe that regions with values $\lesssim 10^2$ K $\kms$ are spatially coincident with the strong continuum emission observed at the centre of CNC (Rebolledo et al., in preparation).

\begin{figure*}
\includegraphics[scale=0.4,angle=-90]{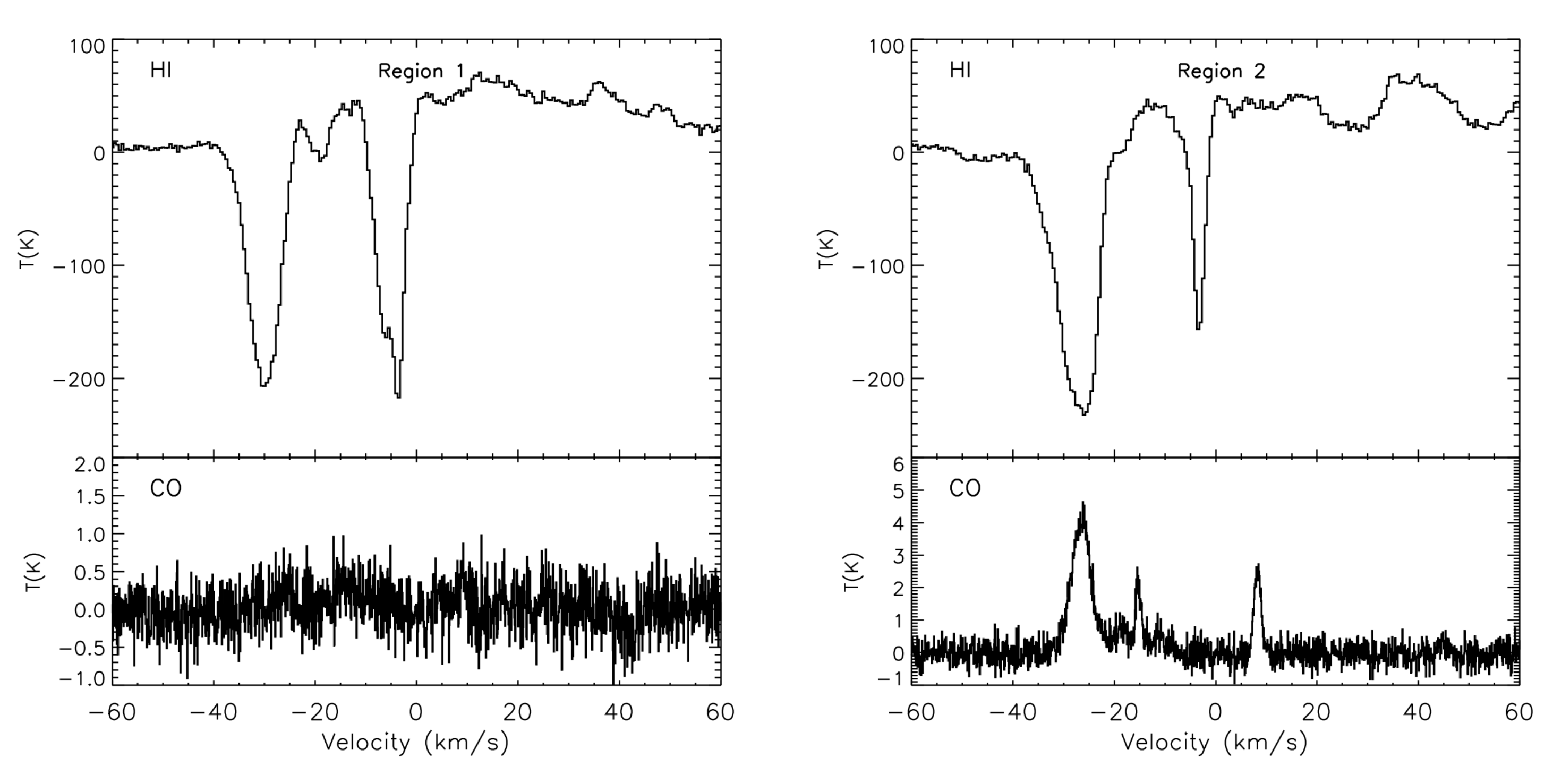} 
\caption{\HI\ absorption features observed across the CNC-Gum 31 complex.  The position of each region is marked in Figure \ref{hi_co}.  In each figure, the top panel shows the \HI\ line profile, while the bottom panel shows the $\co$ spectra from Mopra.  Regions 1 and 2 illustrate the case when a cold cloud of atomic gas observed against strong continuum background emission is seen in absorption.  For Region 1 we do not detect related CO emission, while for Region 2, the cold atomic gas has a molecular counterpart.}
\label{spect_abs}
\end{figure*}

Figure \ref{spect_abs} shows continuum absorption features detected over two regions towards the centre of the CNC.  In these two positions, the continuum emission is so strong that the \HI\ gas is observed in absorption.  The position of each region is shown in Figure \ref{hi_co}.  Region 1 is located in the \hii\ region Car II (\citealt{1968AuJPh..21..881G}) which is being ionised by $\eta$ Carinae and the Trumpler 14 cluster.  Region 2 is located in the  \hii\ region Car I which is being ionised by the star cluster Trumpler 16 (\citealt{1968AuJPh..21..881G}).   For Region 1, we clearly see two strong components at $ -30\ \kms$ and $ -5\ \kms$, and a weaker component at $-20\ \kms$.  Region 2 shows the two stronger velocity components.  The feature at $ -30\ \kms$ is broadened with a negative velocity wing, perhaps suggesting a separate component, and the component at $ -5\ \kms$ is very narrow probably due to less atomic material with that velocity along the line-of-sight.  The differences in line intensity and line width between similar components observed in both Region 1 and 2 are likely to be the result of variation in the continuum emission intensity or the amount of material responsible of the absorption feature.

The central velocities of the \HI\ components in Region 1, $-30$ and $-5\ \kms$, are coincident with the two components detected on the H110$\alpha$ recombination line map reported by \citet{2001MNRAS.327...46B} in Car II.  They detected two velocity components across Car II: one between $-40$ and $-33\ \kms$, and the other between $-18$ and $-4\ \kms$.  Their results suggest that these two components in the ionised gas could be produced by expanding shells in the vicinity of Car II powered by the energy feedback from the nearby star clusters Trumpler 14 and 16.  Thus, the kinematics of the atomic gas are also being affected by the massive star clusters at the heart of the CNC.

\subsubsection{Optical depth and spin temperature profiles}\label{tau-spin}
In order to understand the properties of the atomic gas components of the ISM we have calculated values for the optical depth and spin temperature for some regions in the CNC-Gum 31 complex.  Both quantities can be estimated if there is a strong continuum source observable in the region surveyed.  The technique uses the absorption profile to estimate the optical depth along the line-of-sight to the radio continuum source.  The observed profile may contain both emission and absorption contributions.  Thus, the technique relies on the accurate subtraction of the emission component.  Equation \ref{Tb-all} provides the \HI\ spectrum observed towards the continuum source, the $T_\mathrm{ON}$ spectrum.  If we take the spectrum at a position offset from the continuum source but sufficiently close that we can assume the properties of the gas remain the same, then  $T_\mathrm{OFF}$ is given by Equation \ref{Tb-all} with $T_\mathrm{C}=0$.  The optical depth is then given by

\begin{equation}\label{tau}
\tau=-\mathrm{ln}{\left(\frac{T_\mathrm{ON}-T_\mathrm{OFF}}{T_\mathrm{C}}\right)},
\end{equation}

\noindent and the spin temperature by 

\begin{equation}\label{Tspin}
T_\mathrm{S}=\frac{T_\mathrm{OFF}}{(1-e^{-\tau})}.
\end{equation}

This approach is straightforward, but there are several caveats.  The radio continuum source must be strong ($T_\mathrm{C}$ > $T_\mathrm{S}$) in order to see the \HI\ cloud in absorption.  Also, obtaining an accurate $T_\mathrm{OFF}$ spectrum is problematic if the emission spectrum changes significantly with position. An alternative approach is to use only interferometric observations of the \HI\ absorption profile.  Since the gas producing the emission is generally more diffuse than the colder absorbing gas, observations with an  interferometer will filter out the diffuse emission but will preserve the absorption profile. In this case, the \HI\ optical depth is given by $-\mathrm{ln}{\left(\frac{T_\mathrm{ON}}{T_\mathrm{C}}\right)}$.

\begin{figure}
\centering
\begin{tabular}{c}
\includegraphics[scale=0.42]{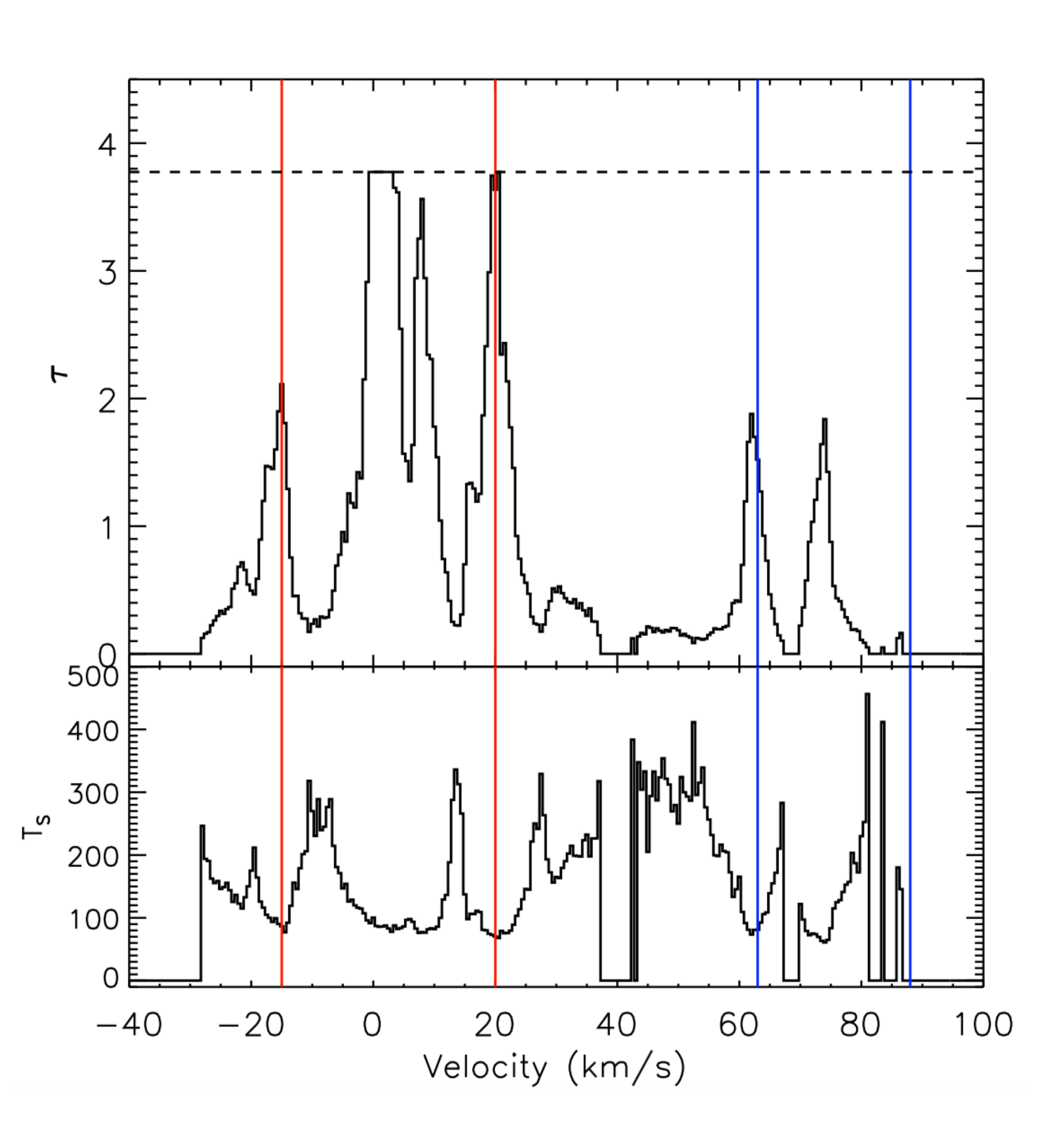} 
\end{tabular}
\caption{{\bf Top}: Optical depth ($\tau$) profile of the \HI\ line along the line-of-sight toward the radio continuum source PMN J1032--5917.  The dashed horizontal line shows the lower limit of $\sim$ 3.8 estimated for $\tau$.  The vertical lines give the velocity range of the spiral arms as in Figure \ref{absoption_compact}.  {\bf Bottom}:  Spin temperature profile of the \HI\ line.  We estimate a $T_\mathrm{S}$ value for each channel with a valid $\tau$ value.  $T_\mathrm{S}$ is $\sim 100$ K or lower for channels with high opacity ($\tau > 2$).} 
\label{tau_Tspin}
\end{figure}

This method of estimating the optical depth of the \HI\ line gives good results for strong continuum sources (\citealt{2015A&A...580A.112B}).  For clouds with $T_\mathrm{S} \sim 100$ K and $\tau \sim 2$, the difference between the two approaches for estimating optical depth is $\sim 20\%$ for continuum sources with $T_\mathrm{C} \sim 1700$ K.  In Figure \ref{absoption_compact} we show the absorption profile toward the radio continuum source PMN J1032--5917 from our interferometric data only.  This continuum source has  $T_\mathrm{C} = 1743$ K, so it is a good candidate for the second simpler approach.  

Because the \HI\ line can be saturated at high opacity, it is necessary to estimate a lower limit for $\tau$, defined by 

\begin{equation}\label{tau_limit}
\tau_\mathrm{lim}=-\mathrm{ln}{\left(\frac{T_\mathrm{thres}}{T_\mathrm{C}}\right)}.
\end{equation}

\noindent where $T_\mathrm{thres}$ is the assumed sensitivity threshold of the \HI\ brightness temperature, set to be $5 \times \sigma_\mathrm{HI}$.  Our ATCA only data have a sensitivity of $\sigma_\mathrm{HI} \sim 8$ K, so we have calculated  $T_\mathrm{thres}=40$ K.  Thus, the lower limit of $\tau$ for saturated lines is 3.8.  

Figure \ref{tau_Tspin} shows the derived \HI\ optical depth from our data.  We find that $\tau$ reaches values larger than 2 for a few clumps of \HI\ gas in the Sagittarius-Carina spiral arm.  The line saturates at velocities $\sim$ 0 $\kms$ and $\sim$ 20 $\kms$.  In the Carina-Gum31 region, $\tau$ reaches a maximum value of $\sim 2.1$ at the velocity $-15\ \kms$.  In the velocity range associated with the Perseus spiral arm two components are identified  with optical depth close to 2 at velocities $\sim 63$ $\kms$ and $\sim 73$ $\kms$.  

Figure \ref{tau_Tspin} also shows the spin temperature derived for each channel with a valid optical depth estimate.  We find $T_\mathrm{S}$ values close to 100 K for the components with $\tau \gtrsim 2$, reflecting the fact that $T_\mathrm{S}$ approaches the brightness temperature for optically thick clouds.  For those channels where $\tau$ is saturated, the estimated $T_\mathrm{S}$ values represent an upper limit, so the actual spin temperature of the absorbing gas is possibly less than 100 K.

\begin{figure*}
\centering
\begin{tabular}{cc}
\includegraphics[scale=0.6]{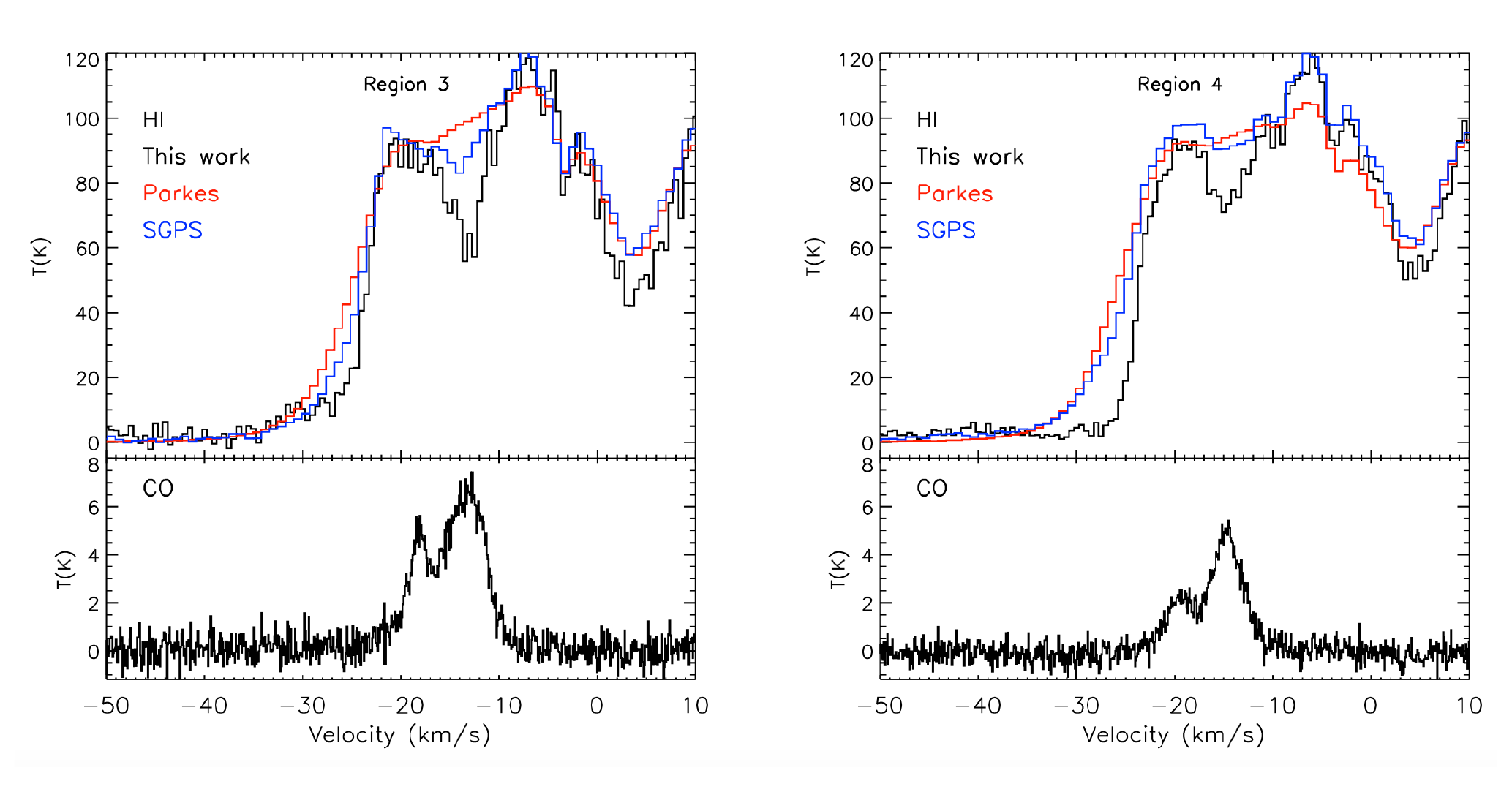} 
\end{tabular}
\caption{HISA features observed in the Gum 31 region.  As in Figure \ref{spect_abs}, the position of each region is marked in Figure \ref{hi_co}.  In each figure, the top panel shows the \HI\ line profile, while the bottom panel shows the $\co$ spectra from Mopra.  In order to emphazise the importance of high spatial resolution in detecting HISA features, we have included the \HI\ spectra from the SGPS survey (blue line) and from only Parkes (red line) data.  The HISA feature is clearly seen in our data, but the SGPS and Parkes profiles do not show the structure because of beam dilution effects.  In both regions we detect associated CO emission.}  
\label{spect_abs_hisa}
\end{figure*}

\subsection{HISA detections}\label{hisa}
HISA detections are an alternative method for tracing the structure of the cold atomic gas, which uses observations towards diffuse warmer atomic gas located in the background.  This approach has been known for many years (\citealt{2000ApJ...540..851G}), but observational constraints have been a limitation.  High spatial resolution observations with an interferometer represent an excellent method to search for HISA.

Figure \ref{spect_abs_hisa} gives two examples of HISA detected in the Gum 31 region.  The location of the regions (Region 3 and 4) used to create the spectra are plotted in Figure \ref{hi_co}.  The areas are 1.5$\times$1.7 arcmin$^2$ and 3.1$\times$2.3 arcmin$^2$ respectively.  For these two examples, we have also included spectra produced from SGPS and Parkes data (\citealt{2005ApJS..158..178M})  to highlight the importance of high spatial resolution in detecting HISA features.  Our data have detected HISA features in both regions, while SGPS and Parkes observations show no clear detection of the self-absorbing cold atomic gas component.  Both HISA features have $\co$ counterparts.  The molecular emission has two clear components, but only one component is clearly distinguishable in the \HI\ spectra.  Assuming that the atomic gas located in this region is cold enough to produce a HISA feature, a possible explanation for this difference could be that the molecular gas with no associated HISA is located in a region where the atomic gas has insufficient column density to produce a detectable absorption feature. 

\begin{figure*}
\centering
\begin{tabular}{c}
\includegraphics[scale=0.3]{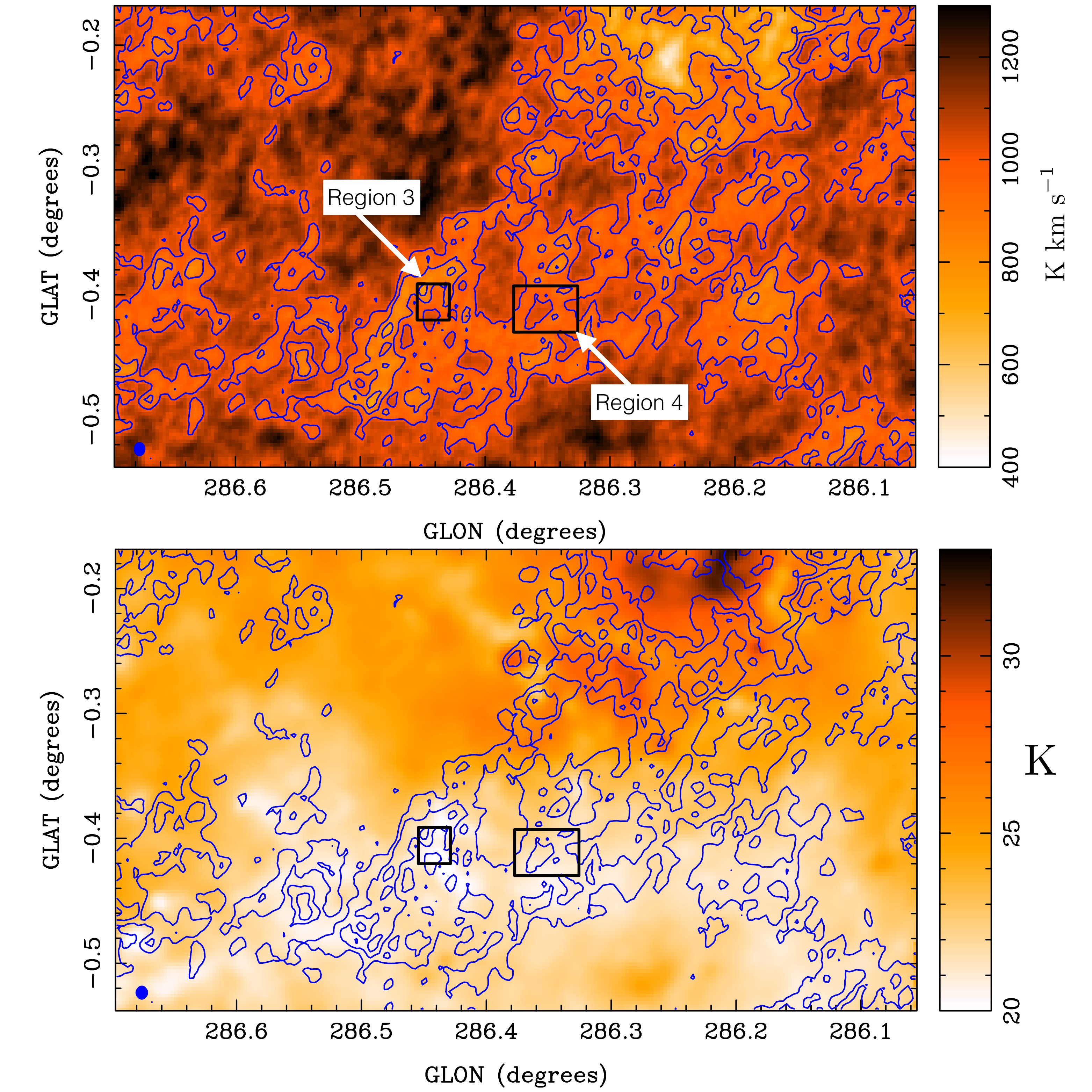} 
\end{tabular}
\caption{{\bf Top}: \HI\ integrated intensity map of the southern part of the Gum 31 region.  We have used a velocity width of 11 $\kms$ centred on the peak velocity of the HISA feature ($-17\ \kms$).  Thus, regions with low integrated intensity might correspond to regions affected by HISA features.  The blue contours are 800, 900 and 1000 K $\kms$.  This particular range has been selected to highlight regions with low integrated intensity (i.\ e.\ affected by HISA).  The black boxes mark the location of the regions used to obtain the averaged spectra shown in Figure \ref{spect_abs_hisa}.  {\bf Bottom}:  Dust temperature map derived in Paper I.  Contours are as in the top panel.  There is good spatial correlation between regions of cold gas as traced by infrared emission (< 22 K) and the distribution of cold atomic gas traced by our radio observations.}  
\label{tdust_hisa}
\end{figure*}

Figure \ref{tdust_hisa} shows a zoomed image covering Regions 3 and 4.  As well as the \HI\ integrated intensity map over the southern part of the Gum 31 region, a map of the dust temperature derived from Herschel infrared maps (from Paper I) is included.  In general, there is a good spatial correlation between regions of cold gas as traced by the infrared emission (< 22 K) and the distribution of cold atomic gas traced by our radio frequency observations.  This spatial agreement between two independent tracers provides further support for the assumption we made in Paper I that the emission in the Herschel infrared maps used to estimate the dust temperatures is in general dominated by the cold component.  

\subsubsection{Optical depth from HISA}
To obtain physical properties from HISA a detailed knowledge of the distribution of the cold and warm components along the line-of-sight is needed to constrain the radiative transfer equation that best describes the feature observed (\citealt{2000ApJ...540..851G} and \citealt{2000ApJ...536..756D}).  Assuming a simple two-component case, we have that a HISA feature is given by

\begin{equation}\label{Tb-hisa-on}
T_\mathrm{HISA-ON}=T_\mathrm{S}(1-e^{-\tau})+ T_\mathrm{W}e^{-\tau},
\end{equation}

\noindent where $T_\mathrm{S}$ and $\tau$ are respectively the spin temperature and optical depth of the absorbing cold gas, and $T_\mathrm{W}$ is the brightness temperature of diffuse warmer gas in the region immediately behind and spatially more extended than the HISA cloud. An expression for the brightness temperature of this warmer gas is be given by

\begin{equation}\label{Tb-hisa-off}
T_\mathrm{HISA-OFF}=T_\mathrm{W}.
\end{equation}

Alternatively, a low spatial resolution observation toward the HISA feature can be used to trace the diffuse emission of the warm atomic gas.  The contribution from the much more compact HISA cloud will be only a small fraction of the total brightness temperature.  Combining Equations \ref{Tb-hisa-on} and \ref{Tb-hisa-off} gives an expression for the optical depth

\begin{equation}\label{eq_tau_hisa}
\tau_\mathrm{HISA}=\mathrm{ln}{\left(\frac{T_\mathrm{HISA-OFF}-T_\mathrm{S}}{T_\mathrm{HISA-ON}-T_\mathrm{S}}\right)}.
\end{equation}

\begin{figure}
\centering
\begin{tabular}{c}
\includegraphics[scale=0.35,angle=-90]{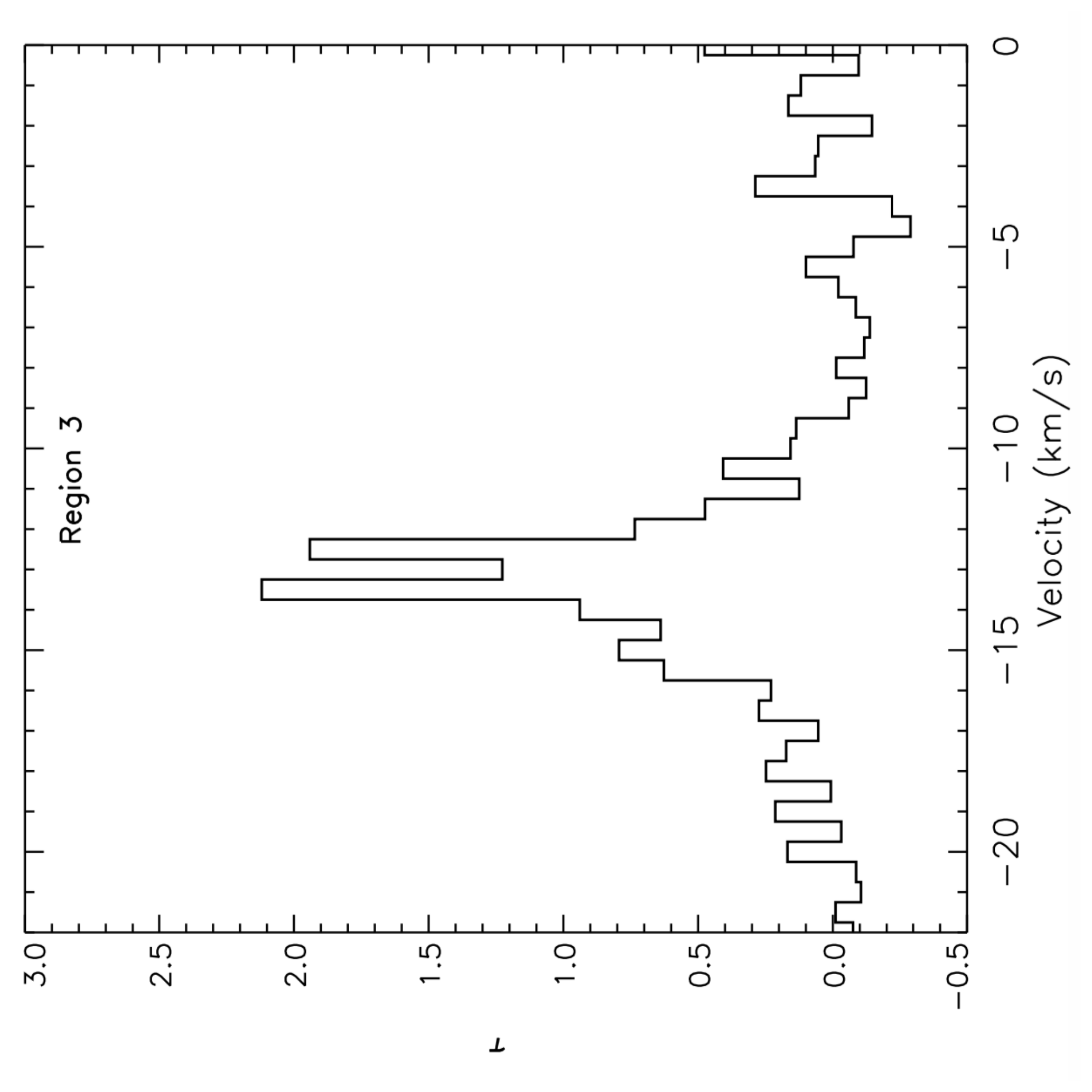} 
\end{tabular}
\caption{Optical depth profile toward the HISA feature identified in Region 3.  We use  $T_\mathrm{S}=50$ K to find optical depth values similar to the values reported in Figure \ref{tau_Tspin}.}  
\label{tau_hisa}
\end{figure}

This equation assumes that all the warm gas is located behind the HISA feature and that there is no contribution from a background continuum radio source. In Figure \ref{tau_hisa} we show the $\tau_\mathrm{HISA}$ profile derived from the line of sight toward the HISA feature in Region 3. To obtain $T_\mathrm{HISA-OFF}$, we have used Parkes observations of the same region (Figure \ref{spect_abs}).  We have assumed $T_\mathrm{S}=50$ K, which gives optical depth estimates consistent with those obtained using the continuum absorption approach (Figure \ref{tau_Tspin}).

\section{Discussion}\label{discuss}

\subsection{Comparison of atomic gas emission and CO}\label{CO_HI_comp}
Our \HI\ map has allowed a direct comparison between the distribution of molecular and atomic gas in the CNC-Gum31 region.  In Figure \ref{hi_co} we compared the integrated intensity maps of the \HI\ and $\co$.  No clear spatial correlation is observed between the two gas phase tracers.  From Section \ref{hi_dist}, the Southern Pillars appear to be co-located with the atomic gas in the 2$\degrees$ long bridge of material extending from the centre of the CNC to negative latitudes (see Figure \ref{cavitities}).  The Southern and Northern Clouds are strongly affected by the diffuse continuum emission of the centre of the CNC, so the \HI\ integrated intensity map is biased towards lower values (see Section \ref{cont_abs}).  Thus, it is not possible to draw any conclusion about the relative distribution of the atomic gas with respect to the molecular gas before the effect of the diffuse continuum is removed from the \HI\ line.  The gas surrounding Gum 31 is located far from the massive star clusters dominating the centre of the CNC.  Consequently, the gas in this part of the complex is less affected by the strong diffuse continuum emission produced by the \hii\ region at the centre of the Carina Nebula.  On the other hand, this region does contain the Gum 31 bubble-shaped \hii\ region dominated by the stellar cluster NGC 3324 (Figure \ref{cont_most}).  With only three identified O-type stars (\citealt{2001A&A...371..107C}; \citealt{2004ApJS..151..103M}), NGC 3324 has fewer massive stars compared to the clusters in the CNC (65 O-type in total, \citealt{2006MNRAS.367..763S}).  The \HI\ integrated intensity map shows a bubble-shaped structure spatially coincident with the position of the \hii\ region.  Although the apparent absence of atomic gas is a plausible explanation for the reduced values of integrated intensity inside the \hii\ region, the presence of the diffuse continuum emission also reduces the \HI\ intensity level in this region. 

\subsection{Atomic gas column density distribution in the Gum 31 region}
\begin{figure}
\centering
\begin{tabular}{c}
\includegraphics[scale=0.4,angle=-90]{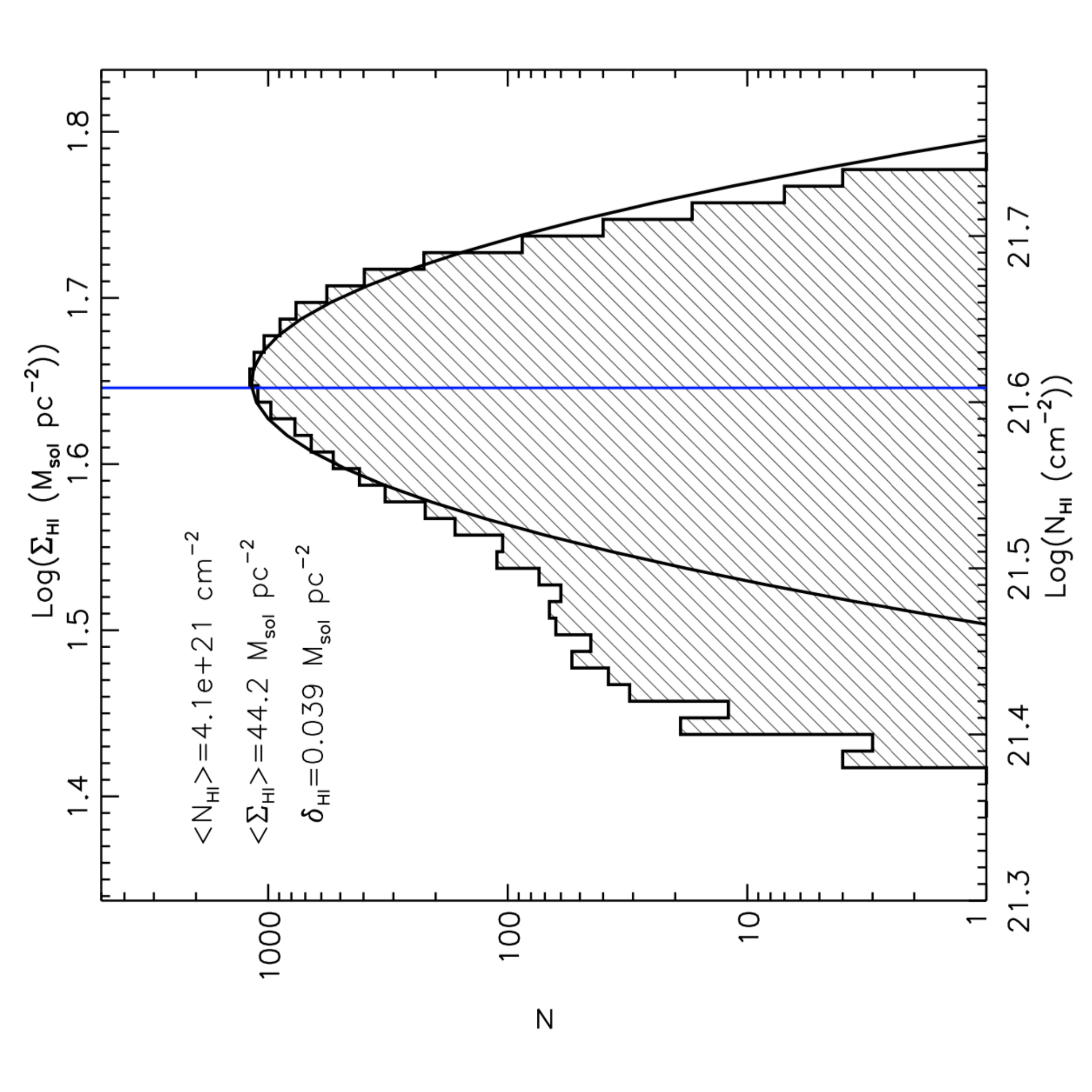} 
\end{tabular}
\caption{Distribution of the \HI\ column density for the Gum 31 region.  We have estimated $N_\mathrm{HI}$ assuming that the \HI\ line emission is optically thin.  The black line shows the lognormal fit to the \HI\ column density distribution.  The fit has been limited to values of  $\mathrm{Log}(N_\mathrm{HI}) > 21.5$.  Estimated from the fitting process, the vertical blue line shows the mean \HI\ column density of  4.1$\times 10^{21}\ \mathrm{cm}^{-2}$, which corresponds to a mass surface density of 44.2 $M_\mathrm{sol}$ $\mathrm{pc}^{-2}$.  The width of the log-normal function is $\delta_\mathrm{HI}=0.04$ $M_\mathrm{sol}$ $\mathrm{pc}^{-2}$.}  
\label{histo_hi}
\end{figure}

In considering the contribution of diffuse continuum emission (Section \ref{cont_diffuse}) to the \HI\ emission line, we need to apply a correction in order to accurately estimate the hydrogen atomic gas column density ($N_\mathrm{HI}$).  The CNC is a very bright thermal region. The impact on the calculation of $N_\mathrm{HI}$ is complicated and will be addressed in a future paper.  On the other hand, the Gum 31 region is not dominated by ionized hydrogen emission and the effect of the diffuse continuum emission on our calculations is minimal.  This region has been used to make preliminary estimates of the atomic gas content.  If we assume that the \HI\ line is optically thin, the \HI\ column density is given by

\begin{equation}\label{NHI_thin}
\frac{N_\mathrm{HI,thin}}{\mathrm{cm^2}}=1.823 \times 10^{18} \int{\frac{T_\mathrm{B,HI}}{\mathrm{K}}\frac{dv}{\kms}}.
\end{equation}

\noindent where the integration is over the range of velocities of gas associated with the target area.  In Paper I, the gas velocities associated with Gum 31  were found to be between $-45\ \kms$ and $0\ \kms$.  Figure \ref{histo_hi} shows the resulting $N_\mathrm{HI}$ distribution for the region.  Following the approach used in Paper I, a log-normal function was fitted to the observed column density distribution in order to investigate if a Gaussian shape properly describes the observed log$(N_\mathrm{HI})$ distribution.  The function used is 

\begin{equation}\label{log-norm}
\mathrm{Num}(\mathrm{pixels})=\mathrm{Num}_\mathrm{peak}\times \exp(-\frac{\log(N_\mathrm{HI})-\log(<N_\mathrm{HI}>)^{2}}{2\times \delta_\mathrm{HI}^2}).
\end{equation}

Figure \ref{histo_hi} shows the functional fit to the data.  The distribution is relatively narrow ($\delta_\mathrm{HI}=0.04$), with a mean value of $<N_\mathrm{HI}>=4.1 \times 10^{21}$ cm$^{-2}$.    The column density distribution  deviates from the model for $\mathrm{log}(N_\mathrm{HI}) < 21.5$.  The excess seen corresponds to one of the bubbles described in Section \ref{hi_dist} and represents only $\sim 5\%$ of the total number of pixels we assigned to the Gum 31 region.  Inside this bubble, the $N_\mathrm{HI}$ values are clearly much reduced.  At high column  densities there is a modest discrepancy between the observed distribution and the log-normal function.  This could  be explained if the \HI\ is becoming optically thick in the denser regions, as was shown by the HISA features discussed in Section \ref{hisa}.

\subsection{Gas mass budget of the Gum 31 region}\label{mass_gas}
Table \ref{mass-budget} summarize the masses for different gas tracers observed in the Gum 31 region.  The masses have been calculated using a mask that excludes pixels with significant continuum emission and pixels with values below the sensitivity limit of each tracer (see Paper I for more details).  For the atomic hydrogen gas, we have used Equation \ref{NHI_thin} to estimate the total mass.  As in Paper I, we use $X_\mathrm{CO}=2\times 10^{20}$ (in units of cm$^{-2}$ (K $\kms$)$^{-1}$), and a gas to dust ratio $R_\mathrm{gd}=100$.   The atomic gas mass, $M_\mathrm{HI}$, is $0.6 \times 10^{5} \Msun$, while the molecular mass is $M_\mathrm{H2}=1.1 \times 10^{5} \Msun$.  The total gas mass obtained from the dust emission is $1.5 \times 10^{5} \Msun$.

Although the sum of the masses estimated from CO and \HI\ are broadly consistent with the gas mass estimated from dust emission, there are intrinsic uncertainties in our calculations.  We have not corrected for opacity effect on the \HI\ line when the atomic mass was calculated (Equation \ref{NHI_thin}).  The impact on mass estimates from the assumption of  an optically thin regime will vary across different regions and there have been studies quantifying these significant differences (\citealt{2015A&A...580A.112B}; \citealt{2015ApJ...809...56L}).  For example, in their study of W43, \citet{2015A&A...580A.112B} find that the atomic mass is a factor of 2.4 larger when opacity corrections are applied.  Consequently, our atomic gas mass estimate would be a lower limit.  The $N_\mathrm{HI}$ and the atomic mass are also highly dependent on the velocity range assumed in Equation \ref{NHI_thin} (\citealt{2012ApJ...748...75L}).  In our calculations we have used the velocity range from $-45\ \kms$ to $0\ \kms$ based on analysis of the CO distribution presented in Paper I.  This could be an overestimate if the selected velocity range includes emission from regions that are not associated with Gum 31.  

\begin{table}
\caption{Gas mass budget for Gum 31.}
\centering
\begin{tabular}{cccc}
\hline\hline
$M$(dust)$^\mathrm{a}$  &  $M_\mathrm{H2}$$^\mathrm{b}$  &   $M_\mathrm{HI}$$^\mathrm{c}$ & $M_\mathrm{H2}$+$M_\mathrm{HI}$ \\
$\Msun$ $\times 10^5$ & $\Msun$ $\times 10^5$  &  $\Msun$ $\times 10^5$ &  $\Msun$  $\times 10^5$    \\ 
\hline
(1.5 $\pm$ 0.1)  & (1.1  $\pm$ 0.1)  & (0.6 $\pm$ 0.1)  & (1.7 $\pm$ 0.1 )   \\
\hline
\multicolumn{4}{l}{{\bf Notes.}} \\
\multicolumn{4}{l}{$^\mathrm{a}$ $M$(dust) is the total gas mass derived from dust emission.}\\
\multicolumn{4}{l}{$^\mathrm{b}$ $M_\mathrm{H2}$ is the molecular gas mass derived from $\co$.}\\
\multicolumn{4}{l}{$^\mathrm{c}$ $M_\mathrm{HI}$ is the atomic gas mass derived from \HI\ 21 cm line.}\\
\end{tabular}
\label{mass-budget}
\end{table}

\begin{figure*}
\centering
\begin{tabular}{c}
\includegraphics[scale=0.4]{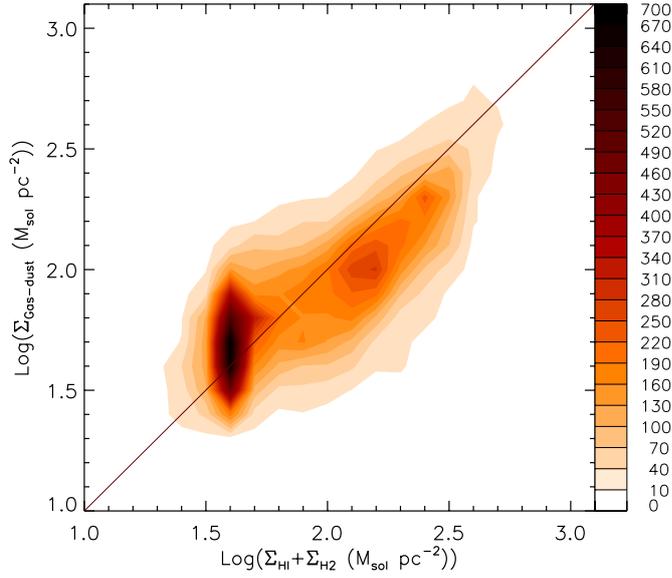} 
\end{tabular}
\caption{Pixel by Pixel comparison between the $\Sigma_\mathrm{gas}$ derived from dust and the $\Sigma_\mathrm{gas}$ estimated by combining the atomic and molecular gas surface density in the Gum 31 region.  The $\Sigma_\mathrm{gas-dust}$ is estimated from the SED fitting of Herschel maps (Paper I), $\Sigma_\mathrm{HI}$ is estimated assuming optically thin emission (Equation \ref{NHI_thin}), and $\Sigma_\mathrm{H2}$ is estimated assuming a $X_\mathrm{CO}=2\times 10^{20} $ cm$^{-2}$ (K $\kms$)$^{-1}$. The colour bar gives the number of pixels analysed.}
\label{comp_dust_hih2}
\end{figure*}

Figure \ref{comp_dust_hih2} shows a pixel by pixel comparison between the gas mass surface density estimated from the dust emission and the gas mass surface density estimated from summing the \HI\ and CO.  In general, there is  a good correlation between the two different tracers of the total gas.  If we look at the region dominated by molecular gas ($\mathrm{Log}(\Sigma_\mathrm{H2}+\Sigma_\mathrm{HI})\gtrsim 2$), we notice that the core of the $\mathrm{Log}(\Sigma_\mathrm{H2}+\Sigma_\mathrm{HI})$ distribution is $\sim$ 0.2 dex (a factor 1.5) larger than the mass gas surface density traced by the dust emission maps ($\mathrm{Log}(\Sigma_\mathrm{Gas-dust})$).  This difference might be explained by an overestimation of the \HI\ mass surface density, or that a larger $X_\mathrm{CO}$ factor should be used to estimate the $\Sigma_\mathrm{H2}$.

\subsection{Atomic to molecular gas transition in the Gum 31 region}
\begin{figure*}
\centering
\begin{tabular}{c}
\includegraphics[scale=0.4]{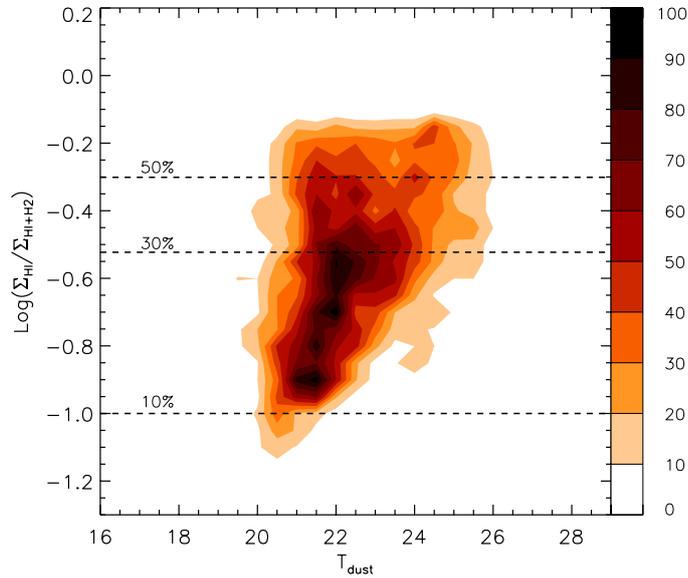} 
\end{tabular}
\caption{Pixel by pixel comparison between the atomic gas fraction ($\Sigma_\mathrm{HI}/\Sigma_\mathrm{HI+H2}$) and the dust temperature in Gum 31.  The horizontal lines shows the percentage of the gas in atomic phase.  In this region, the amount of atomic gas gradually declines as the temperature of the gas becomes colder, reaching a minimum of $\sim10 \%$ of the total for regions where the gas has temperatures $\lesssim 21$ K.  Figure \ref{tdust_hisa} shows the spatial correlation between the cold dust and the regions where we detect HISA features. The colour bar gives the number of pixels analysed.}
\label{comp_tdust_hitohtot}
\end{figure*}

In Figure \ref{tdust_hisa} we show the spatial correlation between the cold \HI\ gas (traced by the HISA features) and the cold dust temperature.  We can identify compact regions of cold gas where the HISA feature is coincident with CO emission, providing evidence for an \HI\ to H$_\mathrm{2}$ transition.  In Figure \ref{comp_tdust_hitohtot} we show a pixel by pixel comparison between the atomic gas fraction ($\Sigma_\mathrm{HI}/\Sigma_\mathrm{HI+H2}$) and the $T_\mathrm{dust}$ for the Gum 31 region.  We observe a gradual reduction of the atomic gas fraction as we move to colder dust temperatures, achieving a minimum of $\sim10 \%$ for the majority of regions with dust temperature $\lesssim 21$ K.  These regions also are positionally coincident with the detection of $\cother$ reported in Paper I.  It is in these regions of cold temperature and high density where the atomic to molecular gas phase transition is likely to be occurring.

\section{Summary}\label{summary}

In this paper we have presented high resolution observations of the \HI\ 21 cm line towards the CNC-Gum 31 molecular complex.  The observing program with the Australia Telescope Compact Array included 11 array configurations to provide a uniformly sampled  $u$-$v$ plane.  We covered $\sim$12 $\deg^2$, centred at $l = 287.5\degrees, b = -1\degrees$, achieving an angular resolution of $\sim 35\arcsec$.  We summarise our main results as follows:

\begin{enumerate}

\item The data cubes reveal complex filamentary structures covering a wide range of velocities.  These filaments are detected in both arm and inter-arm regions, right down to the survey limits. 

\item For the velocity range associated with the CNC-Gum 31 complex, we identify a bridge of atomic gas extending $\sim 2\degrees$ from below the Galactic Plane towards the central part of the CNC.  Several ``bubbles" or cavities are observed in this region (Figure \ref{cavitities}).  The biggest ``bubble" is located below the CNC, which has a bipolar structure produced by the massive star clusters located at the centre of the molecular complex.  Our data suggest that the size of the \HI\ cavity is a factor of 2 larger than previously reported.

\item The \HI\ absorption profile obtained towards the strong radio source PMN J1032--5917 shows the distribution of the cold component of the atomic gas for the velocity range covered by our observations.  Across the Galactic Disk, we find that the most prominent components are associated with the Sagittarius-Carina and Perseus spiral arms.  Inter-arm regions are clearly distinguishable in the spectrum.

\item The strong diffuse continuum emission in the CNC has enabled detection of cold \HI\ components in absorption in some regions of the CNC.  We do not find molecular emission counterparts for all the cold \HI\ components, probably due to differences in the properties of the absorbing material across the molecular complex.

\item The \HI\ absorption profile towards PMN J1032--5917 was used to calculate the optical depth and spin temperature of the cold atomic gas.  We find that the \HI\ is opaque ($\tau \gtrsim$ 2) at several velocities in the Sagittarius-Carina spiral arm.  The spin temperature is $\sim 100$ K in the regions with the highest optical depth, although this value might be less for the saturated components.

\item Assuming an optically thin regime, we estimate the \HI\ column density in the Gum 31 region.  The observed $N_\mathrm{HI}$ distribution is similar to log-normal, except for the highest densities measured. This could be the due to the atomic gas being optically thick in the denser regions.  The atomic mass budget of the Gum 31 region is $0.6 \times 10^{5} \Msun$, which corresponds to $\sim 35 \%$ of the total gas mass as traced by CO+\HI\ maps.  This gas mass is consistent with the total gas mass obtained from the dust emission.  

\item \HI\ self absorption features have revealed the presence of cold atomic gas in  the Gum 31 region.  These HISA features show molecular counterparts and they have a good spatial correlation with the regions of cold dust ($T_\mathrm{dust} \lesssim 21$ K) as traced by the infrared maps.   This implies that regions of cold temperature and high density are where the atomic to molecular gas phase transition is likely to be occurring in this region of the molecular complex.

\end{enumerate}

\section*{Acknowledgements}
The authors are grateful to N. McClure-Griffiths, V. Moss and A. Guzman for useful discussions and significant help with data reduction.   The Australia Telescope Compact Array and the Mopra Telescope are part of the Australia Telescope National Facility, funded by the Australian Government and managed by CSIRO.  Operation of the Mopra  Telescope is made possible by funding from the Australian Research Council and a consortium of institutions through the LIEF grant LE160100094 and the Commonwealth of Australia through CSIRO. DR acknowledges support from the ARC Discovery Project Grant DP130100338, and from CONICYT through project PFB-06 and project Fondecyt 3170568.




\bibliographystyle{mnras}
\bibliography{biblio} 

\begin{thebibliography}{}
\makeatletter
\relax
\def\mn@urlcharsother{\let\do\@makeother \do\$\do\&\do\#\do\^\do\_\do\%\do\~}
\def\mn@doi{\begingroup\mn@urlcharsother \@ifnextchar [ {\mn@doi@}
  {\mn@doi@[]}}
\def\mn@doi@[#1]#2{\def\@tempa{#1}\ifx\@tempa\@empty \href
  {http://dx.doi.org/#2} {doi:#2}\else \href {http://dx.doi.org/#2} {#1}\fi
  \endgroup}
\def\mn@eprint#1#2{\mn@eprint@#1:#2::\@nil}
\def\mn@eprint@arXiv#1{\href {http://arxiv.org/abs/#1} {{\tt arXiv:#1}}}
\def\mn@eprint@dblp#1{\href {http://dblp.uni-trier.de/rec/bibtex/#1.xml}
  {dblp:#1}}
\def\mn@eprint@#1:#2:#3:#4\@nil{\def\@tempa {#1}\def\@tempb {#2}\def\@tempc
  {#3}\ifx \@tempc \@empty \let \@tempc \@tempb \let \@tempb \@tempa \fi \ifx
  \@tempb \@empty \def\@tempb {arXiv}\fi \@ifundefined
  {mn@eprint@\@tempb}{\@tempb:\@tempc}{\expandafter \expandafter \csname
  mn@eprint@\@tempb\endcsname \expandafter{\@tempc}}}

\bibitem[\protect\citeauthoryear{{Bihr} et~al.,}{{Bihr}
  et~al.}{2015}]{2015A&A...580A.112B}
{Bihr} S.,  et~al., 2015, \mn@doi [\aap] {10.1051/0004-6361/201425370}, \href
  {http://adsabs.harvard.edu/abs/2015A%26A...580A.112B} {580, A112}

\bibitem[\protect\citeauthoryear{{Brooks}, {Storey}  \& {Whiteoak}}{{Brooks}
  et~al.}{2001}]{2001MNRAS.327...46B}
{Brooks} K.~J.,  {Storey} J.~W.~V.,   {Whiteoak} J.~B.,  2001, \mn@doi [\mnras]
  {10.1046/j.1365-8711.2001.04590.x}, \href
  {http://adsabs.harvard.edu/abs/2001MNRAS.327...46B} {327, 46}

\bibitem[\protect\citeauthoryear{{Brown}, {Haverkorn}, {Gaensler}, {Taylor},
  {Bizunok}, {McClure-Griffiths}, {Dickey}  \& {Green}}{{Brown}
  et~al.}{2007}]{2007ApJ...663..258B}
{Brown} J.~C.,  {Haverkorn} M.,  {Gaensler} B.~M.,  {Taylor} A.~R.,  {Bizunok}
  N.~S.,  {McClure-Griffiths} N.~M.,  {Dickey} J.~M.,   {Green} A.~J.,  2007,
  \mn@doi [\apj] {10.1086/518499}, \href
  {http://adsabs.harvard.edu/abs/2007ApJ...663..258B} {663, 258}

\bibitem[\protect\citeauthoryear{{Carraro}, {Patat}  \& {Baumgardt}}{{Carraro}
  et~al.}{2001}]{2001A&A...371..107C}
{Carraro} G.,  {Patat} F.,   {Baumgardt} H.,  2001, \mn@doi [\aap]
  {10.1051/0004-6361:20010307}, \href
  {http://adsabs.harvard.edu/abs/2001A%26A...371..107C} {371, 107}

\bibitem[\protect\citeauthoryear{{Clark}}{{Clark}}{1980}]{1980A&A....89..377C}
{Clark} B.~G.,  1980, \aap, \href
  {http://adsabs.harvard.edu/abs/1980A%26A....89..377C} {89, 377}

\bibitem[\protect\citeauthoryear{{Cornwell}, {Braun}  \& {Briggs}}{{Cornwell}
  et~al.}{1999}]{1999ASPC..180..151C}
{Cornwell} T.,  {Braun} R.,   {Briggs} D.~S.,  1999, in {Taylor} G.~B.,
  {Carilli} C.~L.,   {Perley} R.~A.,  eds,  Astronomical Society of the Pacific
  Conference Series Vol. 180, Synthesis Imaging in Radio Astronomy II. p.~151

\bibitem[\protect\citeauthoryear{{Dickey}, {Mebold}, {Stanimirovic}  \&
  {Staveley-Smith}}{{Dickey} et~al.}{2000}]{2000ApJ...536..756D}
{Dickey} J.~M.,  {Mebold} U.,  {Stanimirovic} S.,   {Staveley-Smith} L.,  2000,
  \mn@doi [\apj] {10.1086/308953}, \href
  {http://adsabs.harvard.edu/abs/2000ApJ...536..756D} {536, 756}

\bibitem[\protect\citeauthoryear{{Gaensler}, {McClure-Griffiths}, {Oey},
  {Haverkorn}, {Dickey}  \& {Green}}{{Gaensler}
  et~al.}{2005}]{2005ApJ...620L..95G}
{Gaensler} B.~M.,  {McClure-Griffiths} N.~M.,  {Oey} M.~S.,  {Haverkorn} M.,
  {Dickey} J.~M.,   {Green} A.~J.,  2005, \mn@doi [\apjl] {10.1086/428725},
  \href {http://adsabs.harvard.edu/abs/2005ApJ...620L..95G} {620, L95}

\bibitem[\protect\citeauthoryear{{Gardner} \& {Morimoto}}{{Gardner} \&
  {Morimoto}}{1968}]{1968AuJPh..21..881G}
{Gardner} F.~F.,  {Morimoto} M.,  1968, \mn@doi [Australian Journal of Physics]
  {10.1071/PH680881}, \href {http://adsabs.harvard.edu/abs/1968AuJPh..21..881G}
  {21, 881}

\bibitem[\protect\citeauthoryear{{Gibson}, {Taylor}, {Higgs}  \&
  {Dewdney}}{{Gibson} et~al.}{2000}]{2000ApJ...540..851G}
{Gibson} S.~J.,  {Taylor} A.~R.,  {Higgs} L.~A.,   {Dewdney} P.~E.,  2000,
  \mn@doi [\apj] {10.1086/309364}, \href
  {http://adsabs.harvard.edu/abs/2000ApJ...540..851G} {540, 851}

\bibitem[\protect\citeauthoryear{{Heiles} \& {Troland}}{{Heiles} \&
  {Troland}}{2003a}]{2003ApJS..145..329H}
{Heiles} C.,  {Troland} T.~H.,  2003a, \mn@doi [\apjs] {10.1086/367785}, \href
  {http://adsabs.harvard.edu/abs/2003ApJS..145..329H} {145, 329}

\bibitem[\protect\citeauthoryear{{Heiles} \& {Troland}}{{Heiles} \&
  {Troland}}{2003b}]{2003ApJ...586.1067H}
{Heiles} C.,  {Troland} T.~H.,  2003b, \mn@doi [\apj] {10.1086/367828}, \href
  {http://adsabs.harvard.edu/abs/2003ApJ...586.1067H} {586, 1067}

\bibitem[\protect\citeauthoryear{{H{\"o}gbom}}{{H{\"o}gbom}}{1974}]{1974A&AS...15..417H}
{H{\"o}gbom} J.~A.,  1974, \aaps, \href
  {http://adsabs.harvard.edu/abs/1974A%26AS...15..417H} {15, 417}

\bibitem[\protect\citeauthoryear{{Krumholz}, {McKee}  \&
  {Tumlinson}}{{Krumholz} et~al.}{2009}]{2009ApJ...693..216K}
{Krumholz} M.~R.,  {McKee} C.~F.,   {Tumlinson} J.,  2009, \mn@doi [\apj]
  {10.1088/0004-637X/693/1/216}, \href
  {http://adsabs.harvard.edu/abs/2009ApJ...693..216K} {693, 216}

\bibitem[\protect\citeauthoryear{{Lee} et~al.,}{{Lee}
  et~al.}{2012}]{2012ApJ...748...75L}
{Lee} M.-Y.,  et~al., 2012, \mn@doi [\apj] {10.1088/0004-637X/748/2/75}, \href
  {http://adsabs.harvard.edu/abs/2012ApJ...748...75L} {748, 75}

\bibitem[\protect\citeauthoryear{{Lee}, {Stanimirovi{\'c}}, {Murray}, {Heiles}
  \& {Miller}}{{Lee} et~al.}{2015}]{2015ApJ...809...56L}
{Lee} M.-Y.,  {Stanimirovi{\'c}} S.,  {Murray} C.~E.,  {Heiles} C.,   {Miller}
  J.,  2015, \mn@doi [\apj] {10.1088/0004-637X/809/1/56}, \href
  {http://adsabs.harvard.edu/abs/2015ApJ...809...56L} {809, 56}

\bibitem[\protect\citeauthoryear{{Ma{\'{\i}}z-Apell{\'a}niz}, {Walborn},
  {Galu{\'e}}  \& {Wei}}{{Ma{\'{\i}}z-Apell{\'a}niz}
  et~al.}{2004}]{2004ApJS..151..103M}
{Ma{\'{\i}}z-Apell{\'a}niz} J.,  {Walborn} N.~R.,  {Galu{\'e}} H.~{\'A}.,
  {Wei} L.~H.,  2004, \mn@doi [\apjs] {10.1086/381380}, \href
  {http://adsabs.harvard.edu/abs/2004ApJS..151..103M} {151, 103}

\bibitem[\protect\citeauthoryear{{McClure-Griffiths} \&
  {Dickey}}{{McClure-Griffiths} \& {Dickey}}{2007}]{2007ApJ...671..427M}
{McClure-Griffiths} N.~M.,  {Dickey} J.~M.,  2007, \mn@doi [\apj]
  {10.1086/522297}, \href {http://adsabs.harvard.edu/abs/2007ApJ...671..427M}
  {671, 427}

\bibitem[\protect\citeauthoryear{{McClure-Griffiths}, {Dickey}, {Gaensler},
  {Green}, {Haverkorn}  \& {Strasser}}{{McClure-Griffiths}
  et~al.}{2005}]{2005ApJS..158..178M}
{McClure-Griffiths} N.~M.,  {Dickey} J.~M.,  {Gaensler} B.~M.,  {Green} A.~J.,
  {Haverkorn} M.,   {Strasser} S.,  2005, \mn@doi [\apjs] {10.1086/430114},
  \href {http://adsabs.harvard.edu/abs/2005ApJS..158..178M} {158, 178}

\bibitem[\protect\citeauthoryear{{McKee} \& {Krumholz}}{{McKee} \&
  {Krumholz}}{2010}]{2010ApJ...709..308M}
{McKee} C.~F.,  {Krumholz} M.~R.,  2010, \mn@doi [\apj]
  {10.1088/0004-637X/709/1/308}, \href
  {http://adsabs.harvard.edu/abs/2010ApJ...709..308M} {709, 308}

\bibitem[\protect\citeauthoryear{{Molinari} et~al.,}{{Molinari}
  et~al.}{2010}]{2010PASP..122..314M}
{Molinari} S.,  et~al., 2010, \mn@doi [\pasp] {10.1086/651314}, \href
  {http://adsabs.harvard.edu/abs/2010PASP..122..314M} {122, 314}

\bibitem[\protect\citeauthoryear{{Murphy}, {Mauch}, {Green}, {Hunstead},
  {Piestrzynska}, {Kels}  \& {Sztajer}}{{Murphy}
  et~al.}{2007}]{2007MNRAS.382..382M}
{Murphy} T.,  {Mauch} T.,  {Green} A.,  {Hunstead} R.~W.,  {Piestrzynska} B.,
  {Kels} A.~P.,   {Sztajer} P.,  2007, \mn@doi [\mnras]
  {10.1111/j.1365-2966.2007.12379.x}, \href
  {http://adsabs.harvard.edu/abs/2007MNRAS.382..382M} {382, 382}

\bibitem[\protect\citeauthoryear{{Murray} et~al.,}{{Murray}
  et~al.}{2014}]{2014ApJ...781L..41M}
{Murray} C.~E.,  et~al., 2014, \mn@doi [\apjl] {10.1088/2041-8205/781/2/L41},
  \href {http://adsabs.harvard.edu/abs/2014ApJ...781L..41M} {781, L41}

\bibitem[\protect\citeauthoryear{{Murray} et~al.,}{{Murray}
  et~al.}{2015}]{2015ApJ...804...89M}
{Murray} C.~E.,  et~al., 2015, \mn@doi [\apj] {10.1088/0004-637X/804/2/89},
  \href {http://adsabs.harvard.edu/abs/2015ApJ...804...89M} {804, 89}

\bibitem[\protect\citeauthoryear{{Rebolledo} et~al.,}{{Rebolledo}
  et~al.}{2016}]{2016MNRAS.456.2406R}
{Rebolledo} D.,  et~al., 2016, \mn@doi [\mnras] {10.1093/mnras/stv2776}, \href
  {http://adsabs.harvard.edu/abs/2016MNRAS.456.2406R} {456, 2406}

\bibitem[\protect\citeauthoryear{{Sault}, {Teuben}  \& {Wright}}{{Sault}
  et~al.}{1995}]{1995ASPC...77..433S}
{Sault} R.~J.,  {Teuben} P.~J.,   {Wright} M.~C.~H.,  1995, in {Shaw} R.~A.,
  {Payne} H.~E.,   {Hayes} J.~J.~E.,  eds,  Astronomical Society of the Pacific
  Conference Series Vol. 77, Astronomical Data Analysis Software and Systems
  IV. p.~433 (\mn@eprint {} {astro-ph/0612759})

\bibitem[\protect\citeauthoryear{{Sault}, {Staveley-Smith}  \& {Brouw}}{{Sault}
  et~al.}{1996}]{1996A&AS..120..375S}
{Sault} R.~J.,  {Staveley-Smith} L.,   {Brouw} W.~N.,  1996, \aaps, \href
  {http://adsabs.harvard.edu/abs/1996A%26AS..120..375S} {120, 375}

\bibitem[\protect\citeauthoryear{{Smith}}{{Smith}}{2006}]{2006MNRAS.367..763S}
{Smith} N.,  2006, \mn@doi [\mnras] {10.1111/j.1365-2966.2006.10007.x}, \href
  {http://adsabs.harvard.edu/abs/2006MNRAS.367..763S} {367, 763}

\bibitem[\protect\citeauthoryear{{Smith}, {Egan}, {Carey}, {Price}, {Morse}  \&
  {Price}}{{Smith} et~al.}{2000}]{2000ApJ...532L.145S}
{Smith} N.,  {Egan} M.~P.,  {Carey} S.,  {Price} S.~D.,  {Morse} J.~A.,
  {Price} P.~A.,  2000, \mn@doi [\apjl] {10.1086/312578}, \href
  {http://adsabs.harvard.edu/abs/2000ApJ...532L.145S} {532, L145}

\bibitem[\protect\citeauthoryear{{Smith} et~al.,}{{Smith}
  et~al.}{2010}]{2010MNRAS.406..952S}
{Smith} N.,  et~al., 2010, \mn@doi [\mnras] {10.1111/j.1365-2966.2010.16792.x},
  \href {http://adsabs.harvard.edu/abs/2010MNRAS.406..952S} {406, 952}

\bibitem[\protect\citeauthoryear{{Strasser} \& {Taylor}}{{Strasser} \&
  {Taylor}}{2004}]{2004ApJ...603..560S}
{Strasser} S.,  {Taylor} A.~R.,  2004, \mn@doi [\apj] {10.1086/381674}, \href
  {http://adsabs.harvard.edu/abs/2004ApJ...603..560S} {603, 560}

\bibitem[\protect\citeauthoryear{{Strasser} et~al.,}{{Strasser}
  et~al.}{2007}]{2007AJ....134.2252S}
{Strasser} S.~T.,  et~al., 2007, \mn@doi [\aj] {10.1086/522794}, \href
  {http://adsabs.harvard.edu/abs/2007AJ....134.2252S} {134, 2252}

\bibitem[\protect\citeauthoryear{{Vall{\'e}e}}{{Vall{\'e}e}}{2014}]{2014AJ....148....5V}
{Vall{\'e}e} J.~P.,  2014, \mn@doi [\aj] {10.1088/0004-6256/148/1/5}, \href
  {http://adsabs.harvard.edu/abs/2014AJ....148....5V} {148, 5}

\bibitem[\protect\citeauthoryear{{Walter}, {Brinks}, {de Blok}, {Bigiel},
  {Kennicutt}, {Thornley}  \& {Leroy}}{{Walter}
  et~al.}{2008}]{2008AJ....136.2563W}
{Walter} F.,  {Brinks} E.,  {de Blok} W.~J.~G.,  {Bigiel} F.,  {Kennicutt} Jr.
  R.~C.,  {Thornley} M.~D.,   {Leroy} A.,  2008, \mn@doi [\aj]
  {10.1088/0004-6256/136/6/2563}, \href
  {http://adsabs.harvard.edu/abs/2008AJ....136.2563W} {136, 2563}

\bibitem[\protect\citeauthoryear{{Walterbos} \& {Braun}}{{Walterbos} \&
  {Braun}}{1996}]{1996ASPC..106....1W}
{Walterbos} R.~A.~M.,  {Braun} R.,  1996, in {Skillman} E.~D.,  ed.,
  Astronomical Society of the Pacific Conference Series Vol. 106, The Minnesota
  Lectures on Extragalactic Neutral Hydrogen. p.~1

\bibitem[\protect\citeauthoryear{{Wilson} et~al.,}{{Wilson}
  et~al.}{2011}]{2011MNRAS.416..832W}
{Wilson} W.~E.,  et~al., 2011, \mn@doi [\mnras]
  {10.1111/j.1365-2966.2011.19054.x}, \href
  {http://adsabs.harvard.edu/abs/2011MNRAS.416..832W} {416, 832}

\bibitem[\protect\citeauthoryear{{de Blok}, {Walter}, {Brinks}, {Trachternach},
  {Oh}  \& {Kennicutt}}{{de Blok} et~al.}{2008}]{2008AJ....136.2648D}
{de Blok} W.~J.~G.,  {Walter} F.,  {Brinks} E.,  {Trachternach} C.,  {Oh}
  S.-H.,   {Kennicutt} Jr. R.~C.,  2008, \mn@doi [\aj]
  {10.1088/0004-6256/136/6/2648}, \href
  {http://adsabs.harvard.edu/abs/2008AJ....136.2648D} {136, 2648}

\makeatother
\end{thebibliography}


\bsp	
\label{lastpage}
\end{document}